\documentclass[conference]{IEEEtran}
\IEEEoverridecommandlockouts
\usepackage{cite}
\usepackage{amsmath,amssymb,amsfonts}
\usepackage{algorithmic}
\usepackage{graphicx}
\usepackage{textcomp}
\usepackage{xcolor}
\usepackage{hyperref}
\usepackage{braket}

\usepackage{orcidlink}

\renewcommand{\figureautorefname}{Figure~\negthinspace}

\def\BibTeX{{\rm B\kern-.05em{\sc i\kern-.025em b}\kern-.08em
    T\kern-.1667em\lower.7ex\hbox{E}\kern-.125emX}}
\begin{document}

\title{Recursive QLSTM with Dynamic Variational Quantum Circuit Adaptation \thanks{
The views expressed in this article are those of the authors and do not represent the views of Wells Fargo. This article is for informational purposes only. Nothing contained in this article should be construed as investment advice. Wells Fargo makes no express or implied warranties and expressly disclaims all legal, tax, and accounting implications related to this article.}
}

\author{Samuel Yen-Chi Chen$^1$\orcidlink{0000-0003-0114-4826}, Yifeng Peng$^2$\orcidlink{0009-0007-3306-9417}, Jiun-Cheng Jiang$^6$\orcidlink{0009-0005-1134-4962}, Chun-Hua Lin$^6$\orcidlink{0009-0002-4383-0453}, Kuo-Chung Peng$^6$\orcidlink{0009-0001-8342-2481}\\Junghoon Justin Park$^3$\orcidlink{0000-0001-8982-0387}, Huan-Hsin Tseng$^4$\orcidlink{0000-0001-9544-4226}, Hsin-Yi Lin$^4$\orcidlink{0000-0001-5731-2353}
Kuan-Cheng Chen$^5$\orcidlink{0000-0002-6575-7034}, Chen-Yu Liu$^6$\orcidlink{0000-0002-5437-5188}, Shinjae Yoo$^4$\orcidlink{0000-0003-4378-6448}\\
\small $^1$Wells Fargo
\small $^2$Stevens Institute of Technology
\small $^3$Seoul National University \\
\small $^4$Brookhaven National Laboratory
\small $^5$Imperial College London 
\small $^6$National Taiwan University\\
}

\maketitle

\begin{abstract}
Recent advances in quantum computing and machine learning have motivated the development of quantum models for sequential data processing. In this paper, we propose a Recursive Quantum Long Short-Term Memory model, or Recursive QLSTM, which extends QLSTM through metacore-based recursive constructions. We numerically test the model under different input sequence lengths, metacore designs, and recursive rules, and identify the best-performing architecture among these variants. For this selected model, we further provide theoretical arguments explaining why its recursive structure improves temporal information propagation and enhances learning performance. Our results suggest that Recursive QLSTM offers a flexible and effective framework for quantum recurrent learning over input time series of various lengths.
\end{abstract}


\section{Introduction}
Quantum computing (QC) has the potential to provide computational advantages for certain problem classes \cite{nielsen2010quantum}. In parallel, artificial intelligence and machine learning (AI/ML) have achieved remarkable success across diverse application domains. The emerging field of quantum machine learning (QML) aims to combine these two directions by developing learning models that exploit quantum computational structures.

A central paradigm in QML is the variational quantum algorithm (VQA) \cite{bharti2022noisy,cerezo2021variational}, which forms the basis of many hybrid quantum-classical neural architectures. These variational models have been applied to classification \cite{mitarai2018quantum,schuld2020circuit,hur2022quantum,chen2025validating}, sequence learning \cite{chen2022quantum,li2023qrnn,li2023pqlm}, and reinforcement learning \cite{chen2020QRL,chen2023quantumLSTM_RL,meyer2022survey,chen2025quantum}. Among these applications, sequence learning is particularly important, with use cases ranging from time-series forecasting and speech recognition to natural language processing.

Quantum long short-term memory (QLSTM) networks provide a quantum recurrent framework for modeling temporal data and have demonstrated promising performance in time-series prediction, image generation and language modeling \cite{chen2022quantum,li2023pqlm,di2022dawn,stein2023applying,chu2025lstm,hsu2025quantum,chen2025toward}. However, most existing QLSTM architectures rely on a fixed recurrent structure, where the same quantum module is applied across time steps. This leaves open how quantum recurrent models should be organized to propagate temporal information across input sequences of different fixed lengths, and how alternative recursive structures may affect learning performance.

To address this issue, we study Recursive QLSTM, a metacore-based extension of QLSTM in which temporal information is propagated through explicitly defined recursive rules. This formulation allows different MetaCore designs and recursive rules to be compared within a unified framework, providing a controlled setting for understanding how recursive quantum recurrent structures influence sequence learning. It also motivates a theoretical analysis of the architecture that achieves the strongest empirical performance.

\textbf{Contribution Statement.} The main contributions of this work are threefold. First, we introduce a recursive design framework for Quantum Long Short-Term Memory networks, in which QLSTM dynamics are constructed through reusable metacores and recursively defined information propagation rules. Second, we conduct a systematic numerical study of Recursive QLSTM under different metacore designs, recursive rules, and fixed input sequence lengths, thereby evaluating how architectural and temporal factors affect learning performance. Third, based on the numerical results, we identify the best-performing Recursive QLSTM architecture and provide theoretical arguments explaining why its recursive structure improves temporal information propagation and leads to superior performance.

\section{Related Works}
\label{sec:related_works}

RNNs are a standard tool for nonlinear temporal modeling. Classical NARMA modeling connects recurrent state updates with nonlinear autoregressive moving-average dynamics~\cite{connor1991recurrent}, while LSTM and gated recurrent variants address long-range dependency learning through explicit memory and gating mechanisms~\cite{hochreiter1997long}. Modern surveys on time-series forecasting further indicate that recurrent, convolutional, attention-based, and hybrid predictors mainly differ in how temporal information is stored, transformed, and exposed to the prediction head~\cite{lim2021time}. Our work follows this recurrent time-series modeling setting, but replaces fixed QLSTM gate transformations with dynamically modulated quantum circuit parameters.

Variational quantum algorithms provide the modeling basis for quantum recurrent cells. Near-term quantum learning commonly relies on parameterized quantum circuits (PQCs) optimized by classical routines~\cite{biamonte2017quantum,cerezo2021variational,benedetti2019parameterized,mcclean2018barren,sim2019expressibility,abbas2021power,holmes2022connecting}. Prior QML studies have shown how classical inputs can be encoded into trainable circuits for nonlinear function approximation~\cite{mitarai2018quantum,perezsalinas2020data,schuld2021data,liu2025neural}. These ideas have also been extended to temporal modeling through quantum recurrent neural networks and quantum reservoir computing~\cite{bausch2020recurrent,takaki2021learning,fujii2017harnessing,siemaszko2023rapid,li2023qrnn,araiza2022rydberg,mujal2023timeseriesqrc,hu2024nisqrc,gyurik2026featuremaps,viqueira2025density}. However, such models typically rely on fixed trained parameters, fixed reservoir dynamics, or recurrent quantum-state evolution, rather than gate-wise meta-generated parameter updates at each time step.

QLSTM hybridizes classical LSTM gating with VQCs by replacing the affine transformations in the input, forget, candidate, and output gates with VQC modules~\cite{chen2022quantum}. Related hybrid recurrent models and QLSTM applications have been studied for time-series prediction, renewable-energy forecasting, financial forecasting, dynamical-system prediction, and human activity recognition~\cite{ceschini2022hybrid,khan2024quantum,kea2024hybrid,chen2024quantum,hsu2025federated}. Recent work further explores VQC-instantiated quantum-inspired Kolmogorov--Arnold network (QKAN)-based LSTM models, leading to efficient and scalable quantum-inspired recurrent architectures~\cite{hsu2025qkanlstm,jiang2025qkan}. In contrast to these QLSTM-style models, the proposed recursive QLSTM changes the effective VQC parameters as a function of the recurrent context $z_t=[h_{t-1};c_{t-1};x_t]$, enabling context-adaptive quantum gates during inference.

The proposed recursion is also related to fast-weight and hypernetwork mechanisms, where one model generates or updates another model's parameters from the current context~\cite{schmidhuber1992learning,ba2016using,andrychowicz2016learning,ravi2017optimization,ha2017hypernetworks,munkhdalai2017meta,bertinetto2016learning,jia2016dynamic,yang2019condconv,chen2024qfwp,chen2026qai}. This perspective motivates the MetaCore as a parameter generator for recurrent quantum gates. Our approach is nevertheless distinct from quantum architecture search, which optimizes circuit structures, gate choices, or architecture weights before deployment~\cite{zhang2022differentiable,kuo2021quantum,du2022quantum,wu2023quantumdarts,wang2022quantumnas,chen2024diffqas}. Here, the circuit family is fixed, while the effective VQC parameters are dynamically modulated during sequence processing.

\section{Quantum Neural Networks}\label{Sec: QNN}
\begin{figure}[htbp]
\vskip -0.15in
\centering
\includegraphics[width=1\columnwidth]{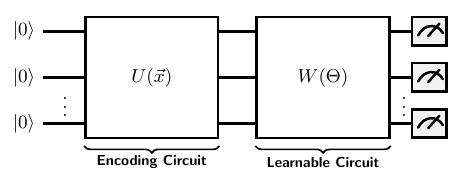}
\caption{\textbf{General variational quantum neural network architecture.} The input is encoded by $U(\mathbf{x})$ and then processed by a learnable quantum circuit $W(\Theta)$.}
\label{fig:QNN_diagram}
\vskip -0.15in
\end{figure}

A quantum neural network (QNN), often formulated as a variational quantum circuit (VQC) or a parameterized quantum circuit (PQC), consists of a data-encoding circuit $U(\vec{x})$, a trainable variational circuit $W(\Theta)$, and a final measurement stage, as illustrated in \figureautorefname{\ref{fig:QNN_diagram}}. For a classical input $\vec{x}$, the quantum state is prepared as $\ket{\Psi}=W(\Theta)U(\vec{x})\ket{0}^{\otimes n},$ where $n$ is the number of qubits and $\Theta$ denotes the trainable circuit parameters. The output is obtained by measuring Hermitian observables $\hat{B}_k$, yielding expectation values
$\langle \hat{B}_k\rangle=\bra{\Psi}\hat{B}_k\ket{\Psi}.$ Thus, the QNN defines a quantum function $\vec{f}(\vec{x};\Theta)=(\langle \hat{B}_1\rangle,\dots,\langle \hat{B}_K\rangle)$.

\begin{figure}[htbp]
\vskip -0.15in
\centering
\includegraphics[width=1\columnwidth]{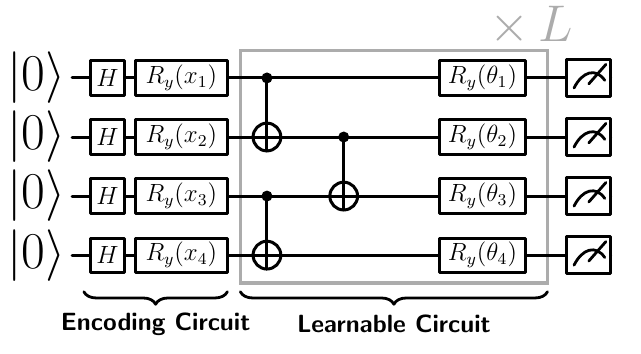}
\caption{\textbf{Detailed VQC structure.} Each qubit is initialized, encoded through input-dependent rotations, and processed by learnable rotation layers repeated over circuit depth $L$.}
\label{fig:VQC_Detail_diagram}
\vskip -0.15in
\end{figure}

In this paper, we employ the VQC/QNN architecture in \figureautorefname{\ref{fig:VQC_Detail_diagram}}, which has been investigated in prior studies \cite{chen2024qfwp}. For an input $u_t$ and VQC parameter matrix $\Theta\in\mathbb{R}^{L\times Q}$, the circuit first applies Hadamard gates and input-encoding rotations, giving $|\psi_t^{(0)}\rangle=\left(\prod_{q=1}^{Q}R_Y(u_{t,q})H_q\right)|0\rangle^{\otimes Q}$. Each variational layer applies a nearest-neighbor CNOT entangling layer $E_Q$ followed by trainable rotations, i.e., $|\psi_t^{(\ell)}\rangle=\left(\prod_{q=1}^{Q}R_Y(\Theta_{\ell q})\right)E_Q|\psi_t^{(\ell-1)}\rangle$ for $\ell=1,\ldots,L$. The VQC output is the vector of Pauli-$Z$ expectation values on the first $H$ qubits, $\mathcal{V}(u_t;\Theta)=[\langle Z_1\rangle_{\psi_t^{(L)}},\ldots,\langle Z_H\rangle_{\psi_t^{(L)}}]^\top\in[-1,1]^H$, where $\langle Z_h\rangle_{\psi_t^{(L)}}=\langle\psi_t^{(L)}|Z_h|\psi_t^{(L)}\rangle$.

\section{Quantum LSTM}\label{Sec: QLSTM}
\begin{figure}[htbp]
\vskip -0.15in
\centering
\includegraphics[width=1\columnwidth]{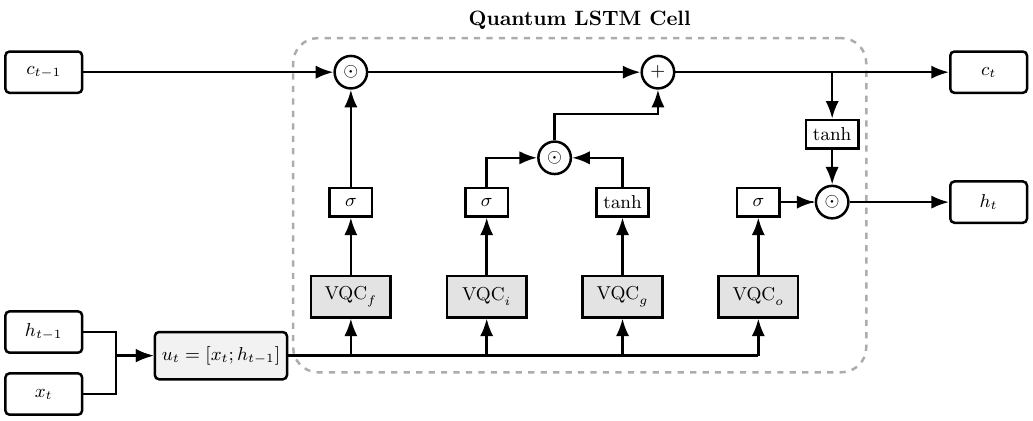}
\caption{\textbf{Quantum LSTM cell.} The input $x_t$ and previous hidden state $h_{t-1}$ are concatenated and processed by four VQC-based gates to update the cell state $c_t$ and hidden state $h_t$.}
\label{fig:QLSTM_main_diagram}
\vskip -0.15in
\end{figure}

Let $x_t\in\mathbb{R}^{D}$ and $h_{t-1},c_{t-1}\in\mathbb{R}^{H}$ denote the
input, previous hidden state, and previous cell state, respectively. The QLSTM
gate input is $u_t=[x_t;h_{t-1}]\in\mathbb{R}^{Q}$ with $Q=D+H$. The VQC uses
$Q$ qubits and $L$ variational layers, and the four gates are indexed by
$a\in\mathcal{A}=\{i,f,g,o\}$ for the input, forget, candidate, and output
gates.
As illustrated in \figureautorefname{\ref{fig:QLSTM_main_diagram}}, the original QLSTM \cite{chen2022quantum} assigns an independent VQC to each LSTM gate. Each
gate has its own trainable parameter matrix
\begin{equation}
    \Theta_i,\Theta_f,\Theta_g,\Theta_o \in \mathbb{R}^{L\times Q}.
\end{equation}
At each time step, the gates are computed as
\begin{align}
    i_t &= \sigma\left(\mathcal{V}(u_t;\Theta_i)\right), \\
    f_t &= \sigma\left(\mathcal{V}(u_t;\Theta_f)\right), \\
    \tilde{c}_t &= \tanh\left(\mathcal{V}(u_t;\Theta_g)\right), \\
    o_t &= \sigma\left(\mathcal{V}(u_t;\Theta_o)\right).
\end{align}
The LSTM state update follows the standard gated recurrence:
\begin{align}
    c_t &= f_t \odot c_{t-1} + i_t \odot \tilde{c}_t, \\
    h_t &= o_t \odot \tanh(c_t), \\
    y_t &= W_y h_t + b_y.
\end{align}
Here $W_y$ and $b_y$ are optional trainable parameters to transform $h_t$ into the prediction $y_t$.

\section{Recursive QLSTM}
The Recursive QLSTM introduces a trainable MetaCore network that dynamically generates or adapts VQC parameters from the current recurrent context. Define
\begin{equation}
    z_t = [h_{t-1}; c_{t-1}; x_t] \in \mathbb{R}^{2H+D}.
\end{equation}
The MetaCore produces four gate-specific parameter updates:
\begin{equation}
    (\Delta^i_t,\Delta^f_t,\Delta^g_t,\Delta^o_t)
    = M_\phi(z_t),
    \qquad
    \Delta^a_t \in \mathbb{R}^{L\times Q}.
\end{equation}
For batch size $B$, the implementation returns
$\Delta^a_t \in \mathbb{R}^{B\times L\times Q}$. The effective VQC parameters
$\Theta^a_t$ are then determined by a recursive rule. The gate equations remain
\begin{align}
    i_t &= \sigma\left(\mathcal{V}(u_t;\Theta^i_t)\right), \\
    f_t &= \sigma\left(\mathcal{V}(u_t;\Theta^f_t)\right), \\
    \tilde{c}_t &= \tanh\left(\mathcal{V}(u_t;\Theta^g_t)\right), \\
    o_t &= \sigma\left(\mathcal{V}(u_t;\Theta^o_t)\right),
\end{align}
followed by the same cell, hidden, and output updates as the original QLSTM.
\begin{figure}[htbp]
\vskip -0.15in
\centering
\includegraphics[width=1\columnwidth]{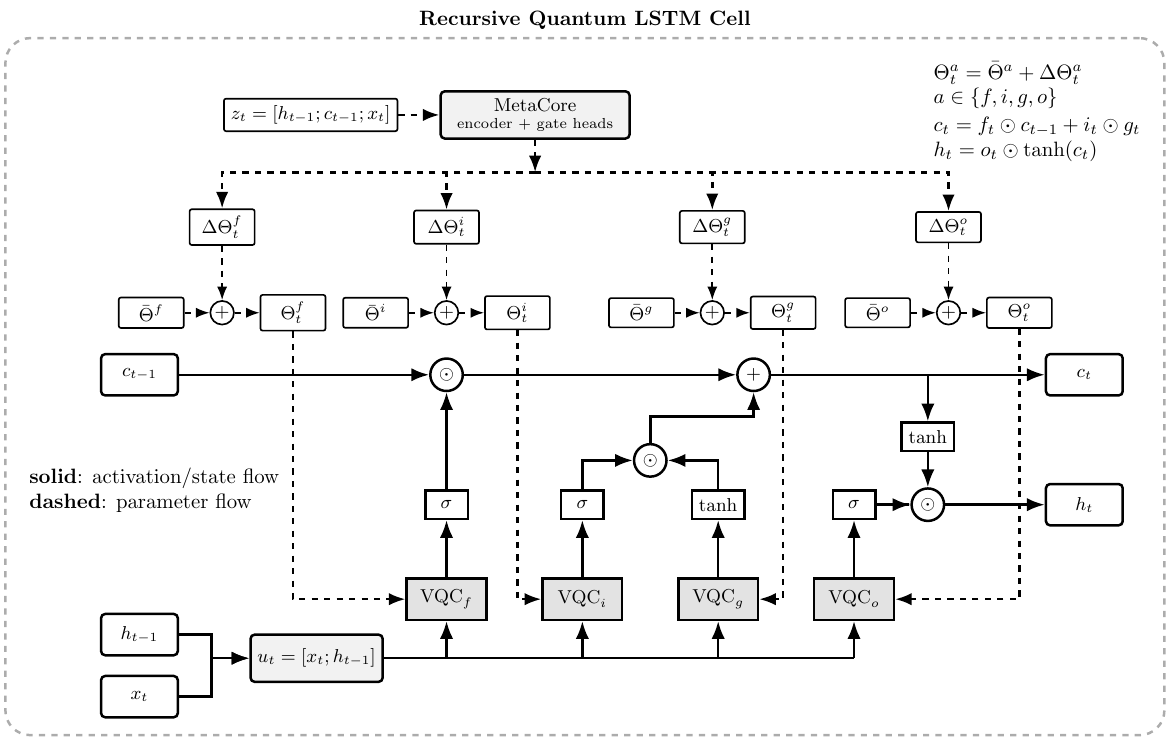}
\caption{\textbf{Recursive Quantum LSTM cell.} The MetaCore generates time-dependent VQC parameter updates from $z_t=[h_{t-1};c_{t-1};x_t]$, enabling adaptive recurrent quantum gates.}
\label{fig:rQLSTM_main_diagram}
\vskip -0.15in
\end{figure}
\subsection{MetaCore Variants}
Let $R=LQ$ and $m$ denote the latent dimension.

\subsubsection{\texttt{SingleNN}}
The single-linear MetaCore directly generates all gate-wise updates as $r_t=Wz_t+b\in\mathbb{R}^{4R}$, followed by $(\Delta^i_t,\Delta^f_t,\Delta^g_t,\Delta^o_t) =\operatorname{reshape}(r_t,4,L,Q)$, making it the simplest recursive parameter generator.

\subsubsection{\texttt{MLP-GELU}}
The MLP-GELU MetaCore increases expressivity by generating updates through a nonlinear embedding:
$e_t=\operatorname{GELU}(W_1z_t+b_1)\in\mathbb{R}^{m}$,
$r_t=W_2e_t+b_2\in\mathbb{R}^{4R}$, and
$(\Delta^i_t,\Delta^f_t,\Delta^g_t,\Delta^o_t)
=\operatorname{reshape}(r_t,4,L,Q)$.

\subsubsection{\texttt{OneNNGate}}
The gate-wise linear MetaCore uses separate linear generators for each gate,
$\Delta^a_t=\operatorname{reshape}(W_a z_t+b_a,L,Q)$, $a\in\{i,f,g,o\}$, thereby decoupling gate-wise parameter updates.

\subsubsection{\texttt{Enc1NNGate}}
\begin{figure}[htbp]
\vskip -0.15in
\centering
\includegraphics[width=1\columnwidth]{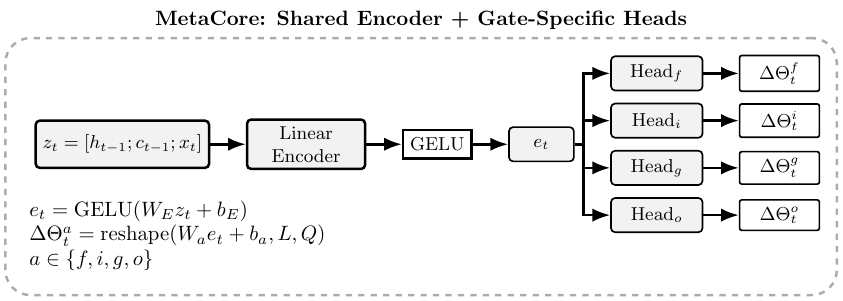}
\caption{\textbf{MetaCore architecture for \texttt{Enc1NNGate}.} A shared encoder extracts the latent representation $e_t$ from $z_t$, while gate-specific heads produce parameter updates for the four QLSTM gates.}
\label{fig:rQLSTM_MetaCore_diagram}
\vskip -0.15in
\end{figure}

The shared-encoder gate-wise MetaCore (shown in \figureautorefname{\ref{fig:rQLSTM_MetaCore_diagram}}) shares a nonlinear embedding
$e_t=\operatorname{GELU}(W_Ez_t+b_E)\in\mathbb{R}^{m}$ across gates, while using gate-specific heads $\Delta^a_t=\operatorname{reshape}(W_ae_t+b_a,L,Q)$, $a\in\{i,f,g,o\}$.

\subsubsection{\texttt{TensorProd}}
The tensor-product MetaCore imposes a rank-one factorization across gate, layer,
and qubit modes by forming
$e_t=W_Ez_t+b_E\in\mathbb{R}^{m}$,
$\lambda_t=W_Le_t+b_L\in\mathbb{R}^{L}$,
$\rho_t=W_Qe_t+b_Q\in\mathbb{R}^{Q}$, and
$\gamma_t=W_Ge_t+b_G\in\mathbb{R}^{4}$, then setting
$\Delta_t=\gamma_t\otimes\lambda_t\otimes\rho_t
\in\mathbb{R}^{4\times L\times Q}$, i.e.,
$\Delta^a_t[\ell,q]=\gamma^a_t\lambda_{t,\ell}\rho_{t,q}$. Thus, each gate-wise update is obtained by fixing the gate index: $\Delta^a_t=\Delta_t[a,:,:]=\gamma^a_t\lambda_t\rho_t^\top \in\mathbb{R}^{L\times Q}$, $a\in\{i,f,g,o\}$.
\subsection{Recursive Rules}
\subsubsection{Base Parameter plus MetaCore Delta (\texttt{base+delta})}
This rule keeps a trainable base parameter matrix for each gate and adds the dynamic MetaCore output, i.e., $\Theta^a_t=\bar{\Theta}^{a}+\Delta^a_t$ for $a\in\{i,f,g,o\}$, where $\bar{\Theta}^{a}\in\mathbb{R}^{L\times Q}$ is shared across time and batch samples. This model can be interpreted as a standard QLSTM whose VQC parameters are adaptively corrected at each time step.

\subsubsection{MetaCore Only (\texttt{meta-only})}
This rule uses the MetaCore output directly as the VQC parameters, i.e., $\Theta^a_t=\Delta^a_t$ for $a\in\{i,f,g,o\}$. In this version, the recurrent context fully determines the gate circuits at each time step, without an additional base VQC parameter.

\subsubsection{MetaCore Delta (\texttt{delta})}
This rule treats the MetaCore output as an increment to the previous VQC parameters, i.e., $\Theta^a_t=\Theta^a_{t-1}+\Delta^a_t$ for $a\in\{i,f,g,o\}$. With zero initialization, this becomes $\Theta^a_t=\sum_{\tau=1}^{t}\Delta^a_\tau$, making the quantum gate parameters themselves recurrent states.
\begin{figure}[htbp]
\vskip -0.15in
\centering
\includegraphics[width=1\columnwidth]{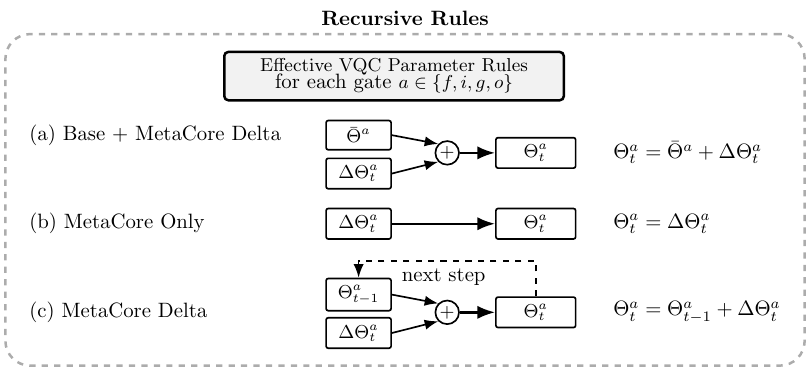}
\caption{\textbf{Recursive parameter update rules for the effective VQC parameters.} The model considers base-plus-delta (\texttt{base+delta}), MetaCore-only (\texttt{meta-only}), and recurrent delta-based (\texttt{delta}) parameter modulation.}
\label{fig:rQLSTM_Recursive_Rules_diagram}
\vskip -0.15in
\end{figure}

\section{Numerical Results and Discussions}
We evaluate the proposed model on five benchmark tasks used in prior studies~\cite{chen2022quantum,chen2024diffqas,chen2024qfwp,chen2022reservoir,chen2025_DiffQAS_QT}: \emph{Damped SHM}, \emph{Bessel function $J_2$}, \emph{Delayed Quantum Control}, \emph{NARMA-5}, and \emph{NARMA-10}. All simulations use \texttt{batch\_size}=4, number of QNN layers $L=5$, \texttt{learning\_rate}=$10^{-3}$, Adam, and random seeds $0$, $1$, and $2$.
Numerical results are organized by task, with three complementary views per dataset. First, epoch-wise prediction snapshots at sequence length 16 provide a trajectory-level comparison between the baseline QLSTM and a representative recursive model. Second, phase-0 train/test convergence curves at the same sequence length screen all candidate recursive variants. Third, phase-1 results across sequence lengths 4, 8, 16, 32, and 64 summarize the retained subset of variants. All curves are averaged over three random seeds, with shaded regions denoting the corresponding standard deviation.

Phase 0 aims to identify a compact set of representative and competitive recursive designs prior to the full sequence-length study. Based on its results, we retain two MetaCore choices, \texttt{Enc1NNGate} and \texttt{MLP-GELU}, each combined with the three recursive rules \texttt{base+delta}, \texttt{delta}, and \texttt{meta-only} for the subsequent phase-1 experiments.
To complement the final test loss, which captures only end-point accuracy, we further report two optimization-oriented metrics on the seed-averaged test-loss curve: AUC@20 and $t95$. Let $\ell_e$ be the seed-averaged test loss at logged epoch $e$, with $\ell_1$ and $\ell_T$ denoting the first and last logged values. \textbf{AUC@20} measures early-stage optimization by the trapezoidal area over the first $20$ logged points, $\mathrm{AUC@20}=\sum_{k=1}^{19}(\ell_k+\ell_{k+1})(e_{k+1}-e_k)/2$, where smaller values indicate faster early error reduction. \textbf{$t95$} denotes the first epoch at which $95\%$ of the total improvement $\Delta=\ell_1-\ell_T$ has been achieved, i.e., $t95=\min\{e:\ell_e\le \ell_T+0.05\Delta\}$; if $\Delta\le0$ or the threshold is never reached, $t95$ is set to the final epoch. Thus, a smaller $t95$ indicates earlier convergence to a near-final loss level.
\begin{figure}[htbp]
\vskip -0.15in
\centering
\includegraphics[width=1\columnwidth]{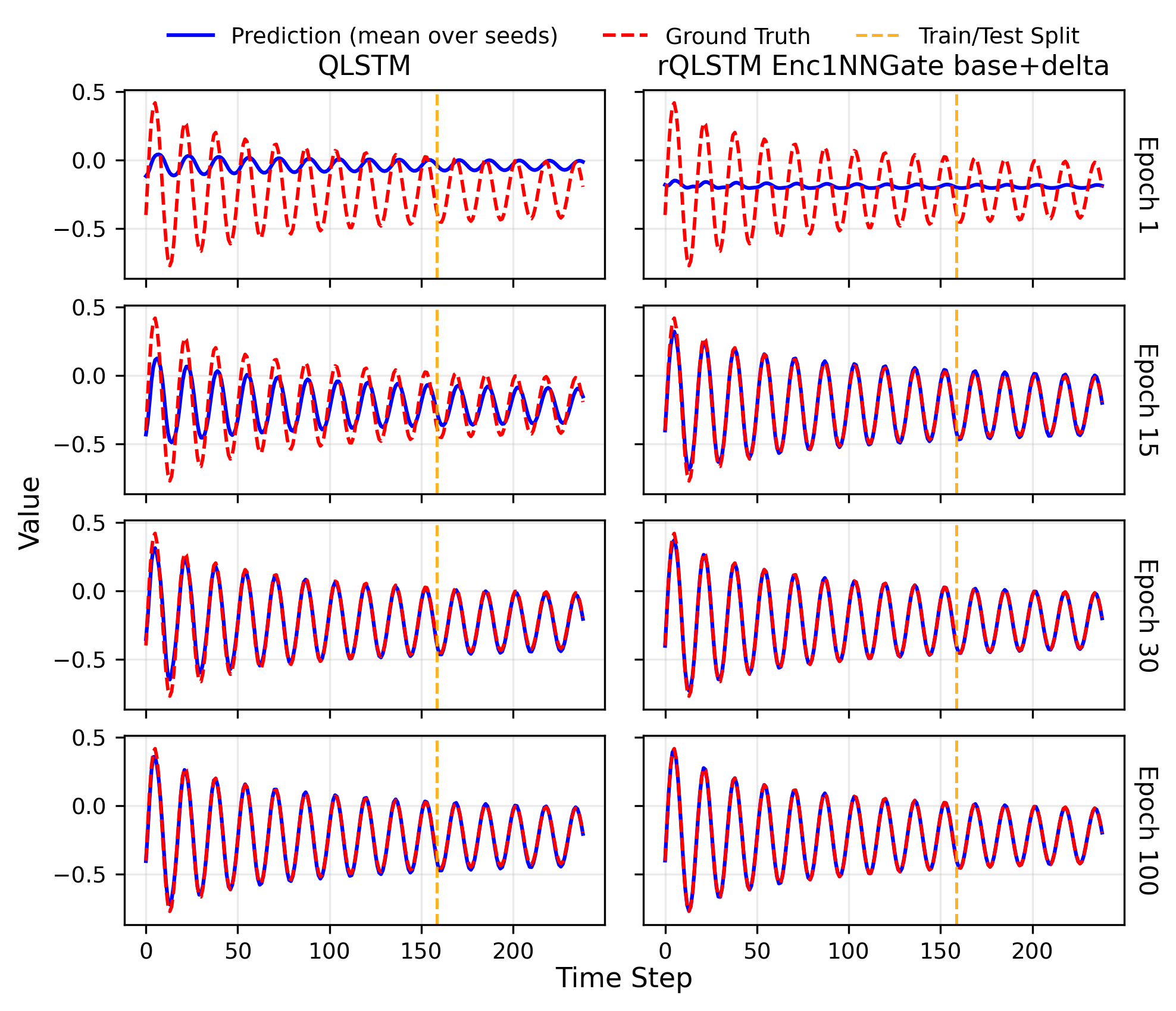}
\caption{\textbf{Epoch-wise prediction comparison between QLSTM and rQLSTM Enc1NNGate base+delta on the \texttt{bessel\_j2} dataset under the \texttt{seq\_len}=16 setting.} Rows show selected training epochs. Blue curves denote the mean prediction over seeds, red dashed curves denote the ground truth, and the orange dashed line marks the train/test split.}
\label{fig:bessel_j2_rollout}
\vskip -0.15in
\end{figure}
\paragraph{Bessel-$J_2$.}
We begin with the \texttt{bessel\_j2} task. 
\figureautorefname{\ref{fig:bessel_j2_rollout}} compares QLSTM with \texttt{rQLSTM Enc1NNGate base+delta} at sequence length $16$. At epoch 1, both models fail to reproduce the decaying oscillatory structure. By epoch 15, the recursive model already matches the waveform across the train and test regions, whereas QLSTM still shows visible amplitude mismatch. The gap narrows at epoch 30, and both achieve high accuracy by epoch 100, indicating that recursion mainly accelerates recovery of the underlying dynamics.
\begin{figure}[htbp]
\vskip -0.15in
\centering
\includegraphics[width=1\columnwidth]{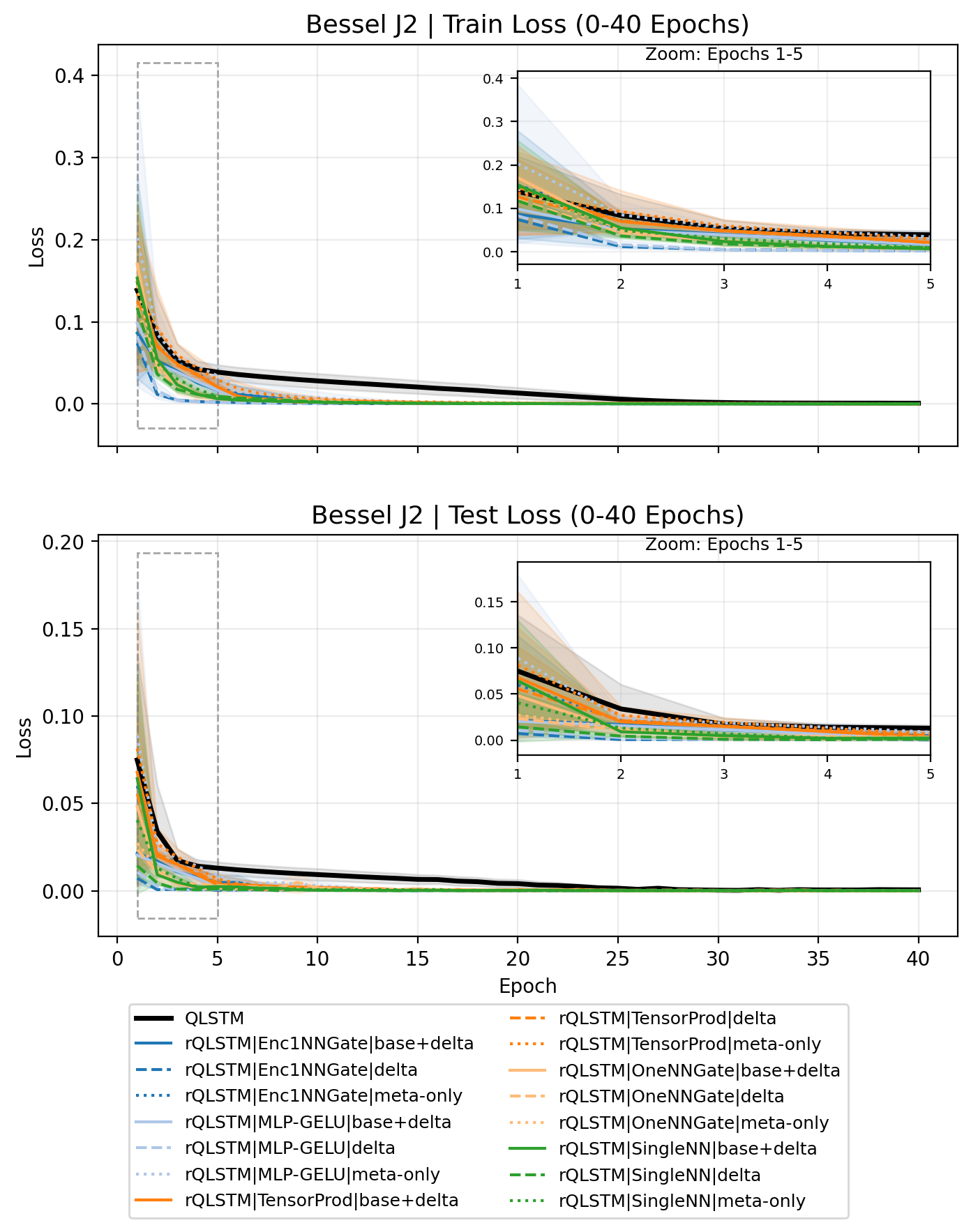}
\caption{\textbf{Train and test loss convergence comparison on the \texttt{bessel\_j2} dataset with \texttt{seq\_len}=16.} We compare the baseline QLSTM against all rQLSTM variants with different MetaCore designs and recursive rules. Curves show the mean over seeds with shaded regions indicating standard deviation, and the inset zooms into the first five epochs.}
\label{fig:bessel_j2_phase_0_convergence_compare}
\vskip -0.15in
\end{figure}

\figureautorefname{\ref{fig:bessel_j2_phase_0_convergence_compare}} reports the phase-0 train/test convergence results at sequence length 16, comparing all candidate recursive variants against the QLSTM baseline. A consistent trend emerges in both training and testing loss: nearly all recursive variants converge substantially faster than the non-recursive QLSTM, with the largest advantage appearing in the first few epochs, as highlighted by the inset plots over epochs 1--5. Differences among the recursive variants themselves are relatively modest compared with the much larger gap between the recursive family and the baseline. This indicates that, for \texttt{bessel\_j2}, the primary optimization gain stems from recursion itself, while the specific MetaCore and recursive-rule choices yield only finer-grained differences in early-stage convergence, thereby supporting the use of a compact screening step before the full sequence-length study.
\begin{figure}[htbp]
\vskip -0.15in
\centering
\includegraphics[width=1\columnwidth]{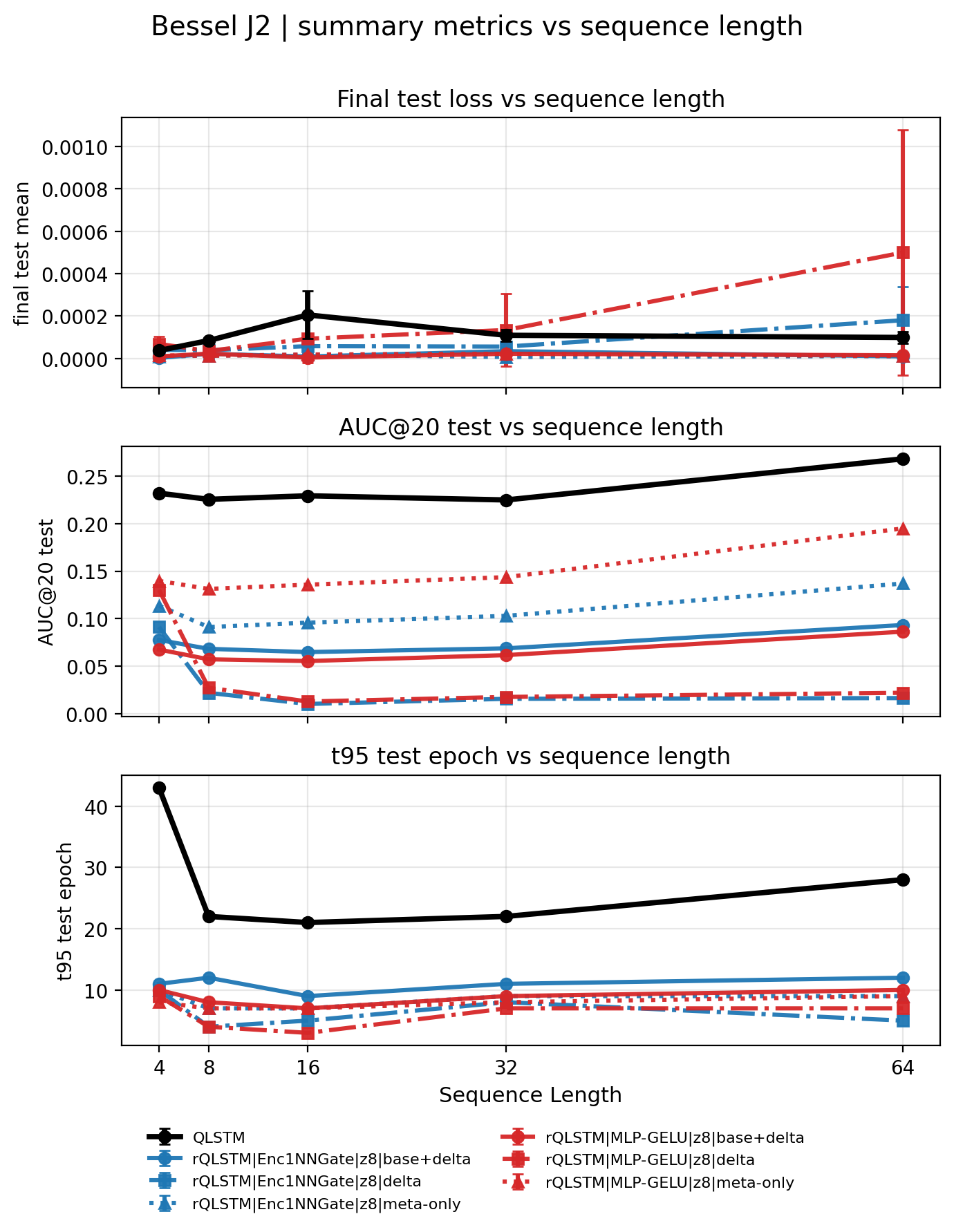}
\caption{\textbf{Summary performance comparison versus sequence length on the \texttt{bessel\_j2} dataset for sequence lengths 4, 8, 16, 32, and 64.} We compare the baseline QLSTM with selected rQLSTM variants under different MetaCore designs and recursive update rules. The three panels report final test loss, AUC@20 test loss, and $t95$ test epoch, respectively.}
\label{fig:bessel_j2_summary_over_seq_len}
\vskip -0.15in
\end{figure}

\figureautorefname{\ref{fig:bessel_j2_summary_over_seq_len}} summarizes the phase-1 results across sequence lengths 4, 8, 16, 32, and 64 for the retained variants, with the three panels reporting final test loss, AUC@20 test loss, and $t95$ test epoch, respectively. 
The overall pattern favors the recursive family: while final test losses are comparable to or lower than those of QLSTM, the improvement is especially pronounced in optimization-oriented metrics, where most recursive variants achieve lower AUC@20 and markedly smaller $t95$, indicating faster error reduction and earlier convergence across temporal horizons. 
Among the retained designs, the \texttt{base+delta} variants are particularly stable across sequence lengths, whereas certain \texttt{delta} and \texttt{meta-only} settings remain competitive but exhibit slightly larger variability at longer horizons. 
Overall, the \texttt{bessel\_j2} results demonstrate that recursive QLSTM variants preserve strong final predictive accuracy while providing a consistent advantage in training efficiency over the standard QLSTM baseline.

\begin{figure}[htbp]
\vskip -0.15in
\centering
\includegraphics[width=1\columnwidth]{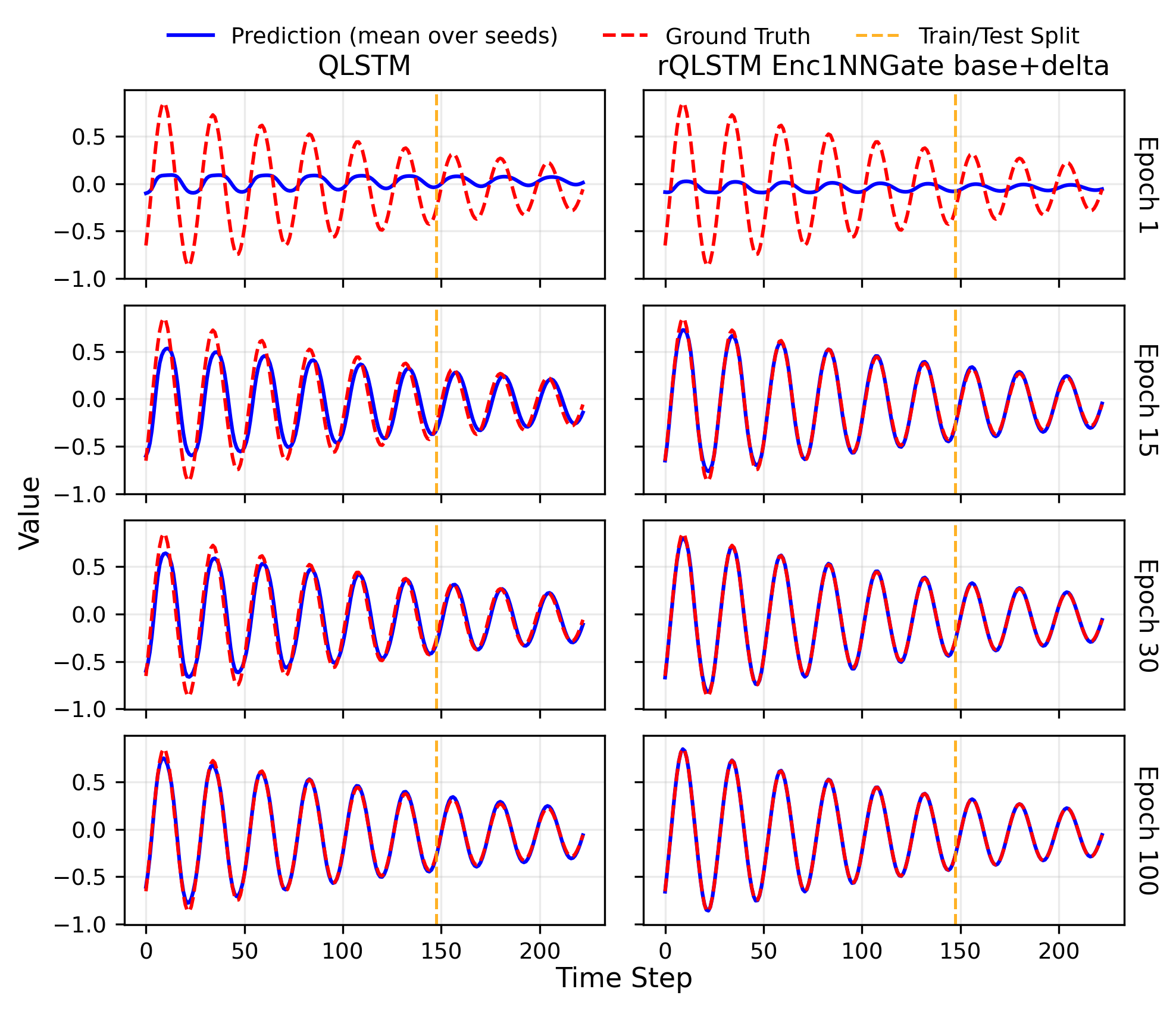}
\caption{\textbf{Epoch-wise prediction comparison between QLSTM and rQLSTM Enc1NNGate base+delta on the \texttt{damped\_shm} dataset under the \texttt{seq\_len}=16 setting.} Rows show selected training epochs. Blue curves denote the mean prediction over seeds, red dashed curves denote the ground truth, and the orange dashed line marks the train/test split.}
\label{fig:damped_shm_rollout}
\vskip -0.15in
\end{figure}
\paragraph{Damped-SHM.}
We next consider the \texttt{damped\_shm} task. \figureautorefname{\ref{fig:damped_shm_rollout}} presents the epoch-wise prediction snapshots at sequence length $16$, comparing the standard QLSTM with the representative recursive model \texttt{rQLSTM Enc1NNGate base+delta}. As in the \texttt{bessel\_j2} case, both models are inaccurate at the beginning of training, but the recursive model approaches the target waveform substantially faster. By epoch 15, it already captures the damped oscillatory structure with high fidelity across both the training and testing regions, whereas the baseline QLSTM still exhibits visible mismatch in amplitude and waveform alignment. The difference gradually narrows as training proceeds, and by epoch 100 both models achieve accurate predictions. This again indicates that the recursive formulation primarily improves learning efficiency by accelerating the recovery of the underlying temporal dynamics.
\begin{figure}[htbp]
\vskip -0.12in
\centering
\includegraphics[width=1\columnwidth]{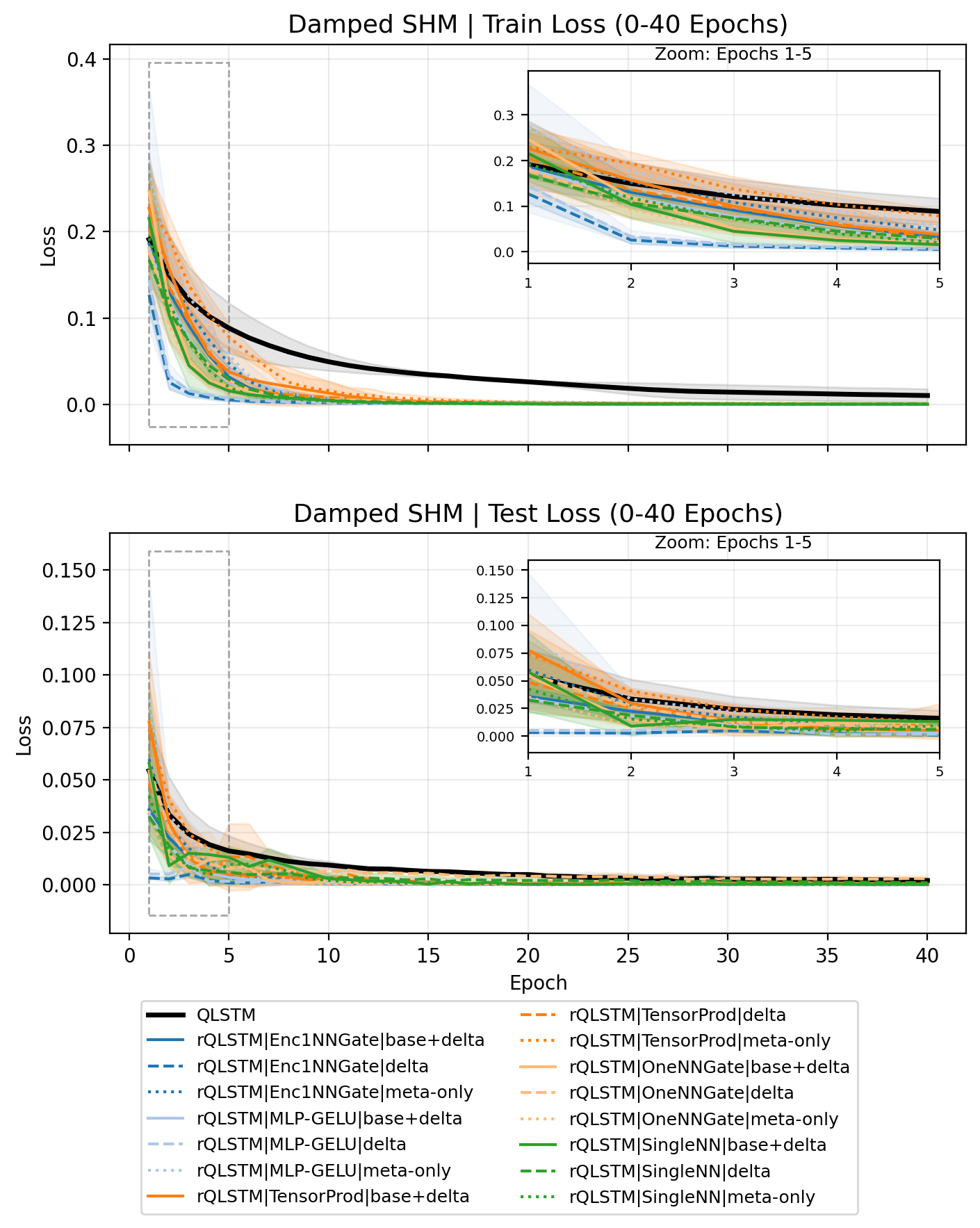}
\caption{\textbf{Train and test loss convergence comparison on the \texttt{damped\_shm} dataset with \texttt{seq\_len}=16.} We compare the baseline QLSTM against all rQLSTM variants with different MetaCore designs and recursive rules. Curves show the mean over seeds with shaded regions indicating standard deviation, and the inset zooms into the first five epochs.}
\label{fig:damped_shm_phase_0_convergence_compare}
\vskip -0.12in
\end{figure}

\figureautorefname{\ref{fig:damped_shm_phase_0_convergence_compare}} reports the phase-0 convergence results at sequence length $16$, comparing all recursive variants against the QLSTM baseline. 
The overall trend is highly consistent with the previous task: in both training and testing loss, most recursive variants converge substantially faster than the non-recursive QLSTM, with the advantage especially evident during the first few epochs, as highlighted by the inset plots. 
Although small differences can be observed among MetaCore and recursive-rule combinations, they remain secondary compared with the much larger gap between the recursive family and the baseline, further supporting the role of phase 0 as a screening stage prior to the full sequence-length study.
\begin{figure}[htbp]
\vskip -0.12in
\centering
\includegraphics[width=1\columnwidth]{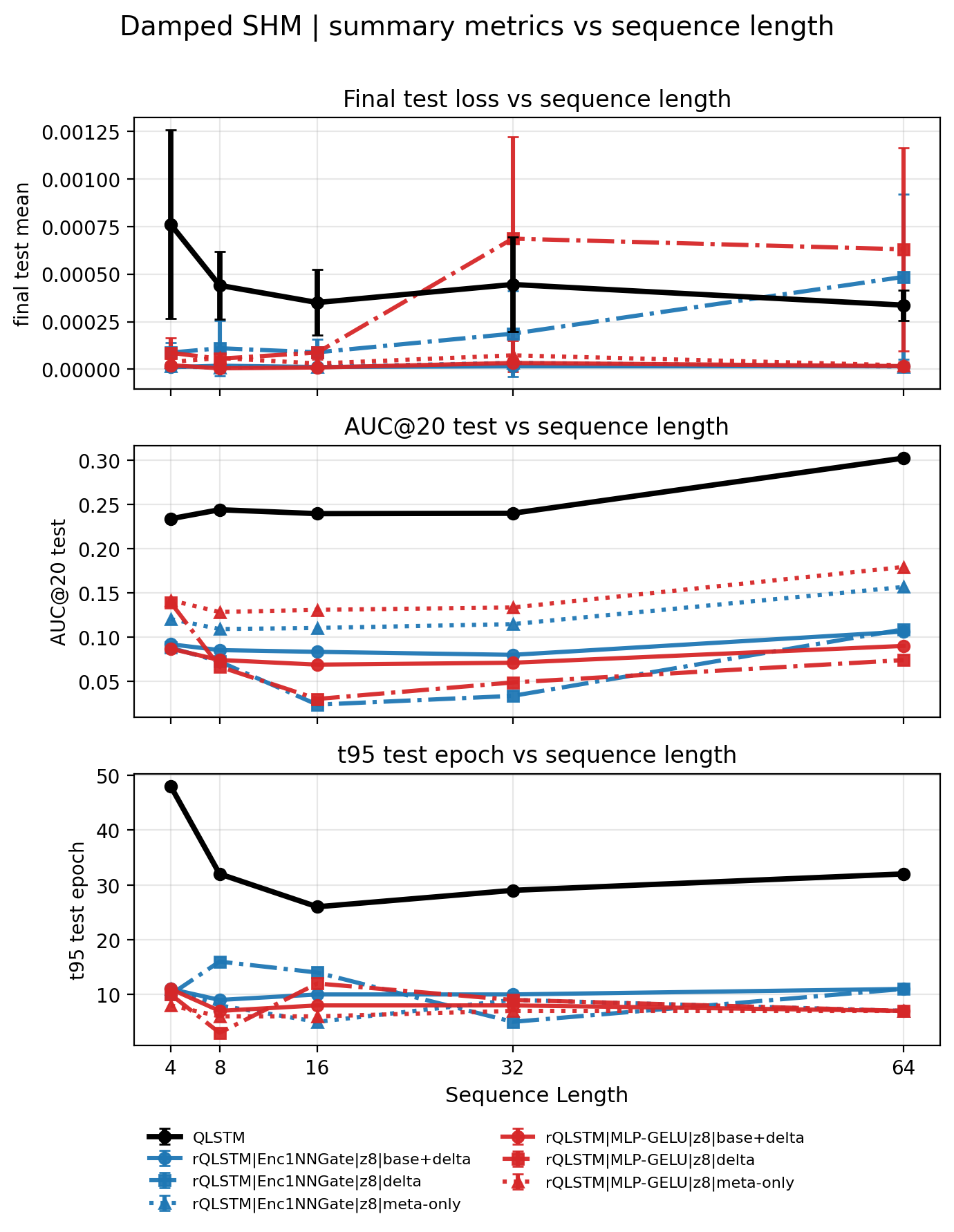}
\caption{\textbf{Summary performance comparison versus sequence length on the \texttt{damped\_shm} dataset for sequence lengths 4, 8, 16, 32, and 64.} We compare the baseline QLSTM with selected rQLSTM variants under different MetaCore designs and recursive update rules. The three panels report final test loss, AUC@20 test loss, and $t95$ test epoch, respectively.}
\label{fig:damped_shm_summary_over_seq_len}
\vskip -0.1in
\end{figure}

\figureautorefname{\ref{fig:damped_shm_summary_over_seq_len}} summarizes the phase-1 results across sequence lengths $4, 8, 16, 32,$ and $64$ for the retained variants, with the three panels reporting final test loss, AUC@20 test loss, and $t95$ test epoch, respectively. The recursive models again show their clearest advantage in optimization-related metrics: compared with QLSTM, they achieve much lower AUC@20 and markedly smaller $t95$, indicating faster error reduction and earlier convergence over different temporal horizons. Several recursive variants also remain competitive or superior in final test loss, though some settings exhibit more noticeable variability as the sequence length increases. Overall, the \texttt{damped\_shm} results reinforce the conclusion that recursive QLSTM designs offer a consistent convergence advantage while preserving strong predictive performance.
\begin{figure}[htbp]
\vskip -0.1in
\centering
\includegraphics[width=1\columnwidth]{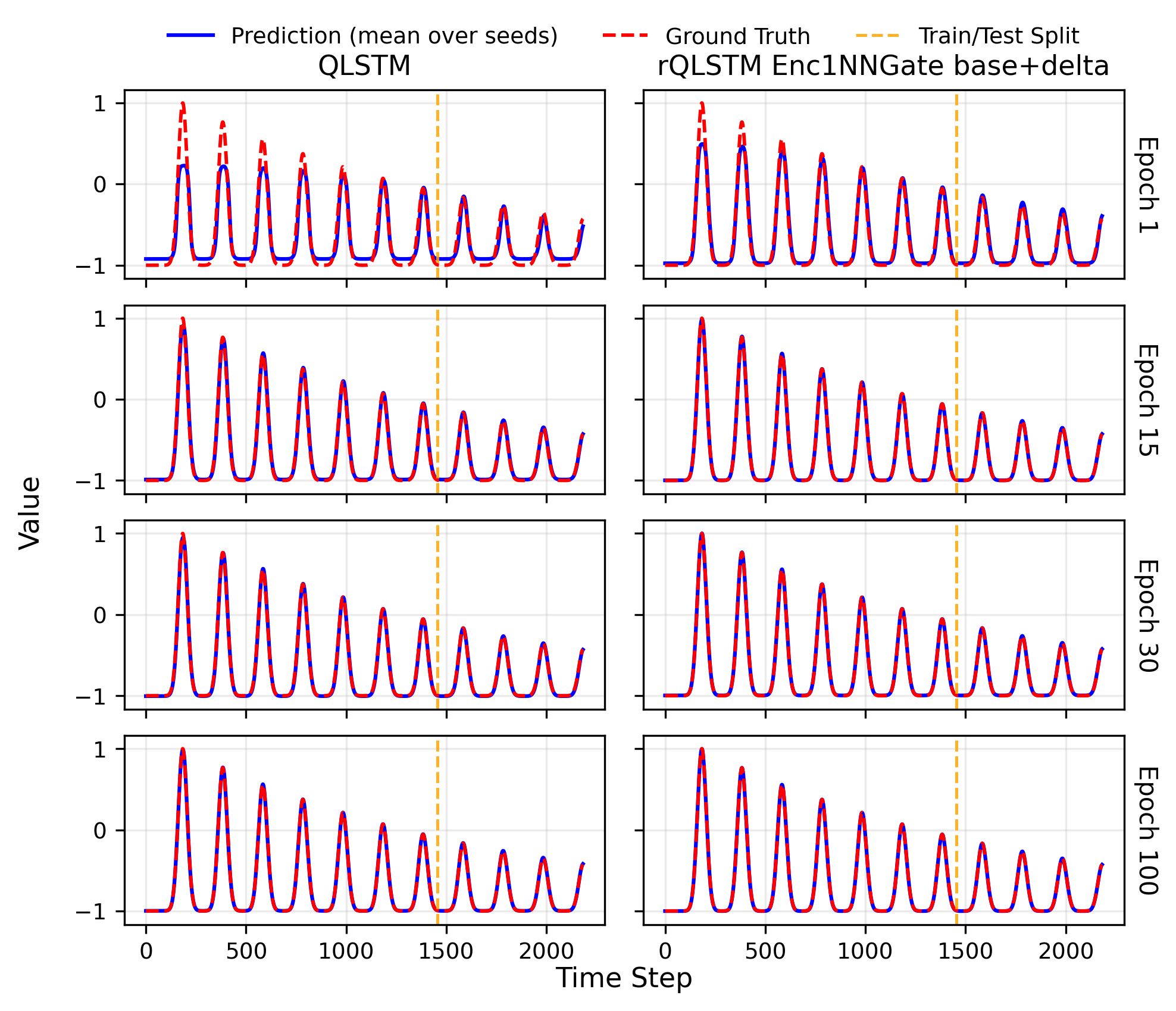}
\caption{\textbf{Epoch-wise prediction comparison between QLSTM and rQLSTM Enc1NNGate base+delta on the \texttt{delayed\_quantum\_control} dataset under the \texttt{seq\_len}=16 setting.} Rows show selected training epochs. Blue curves denote the mean prediction over seeds, red dashed curves denote the ground truth, and the orange dashed line marks the train/test split.}
\label{fig:delayed_quantum_control_rollout}
\vskip -0.1in
\end{figure}
\paragraph{Delayed Quantum Control.}
We next evaluate the \texttt{delayed\_quantum\_control} task. \figureautorefname{\ref{fig:delayed_quantum_control_rollout}} presents the epoch-wise prediction snapshots at sequence length $16$, comparing the standard QLSTM with the representative recursive model \texttt{rQLSTM Enc1NNGate base+delta}. 
At the start of training, both models are underfitted and fail to fully reproduce the target trajectory. 
The recursive model approaches the target dynamics slightly earlier, and by epoch 15 both models already recover the waveform reasonably well across the training and testing regions. 
From epoch 30 onward, their predictions become nearly indistinguishable. For this task, the main benefit of recursion is therefore reflected in earlier fitting rather than in large differences in final trajectory quality.
\begin{figure}[htbp]
\vskip -0.1in
\centering
\includegraphics[width=1\columnwidth]{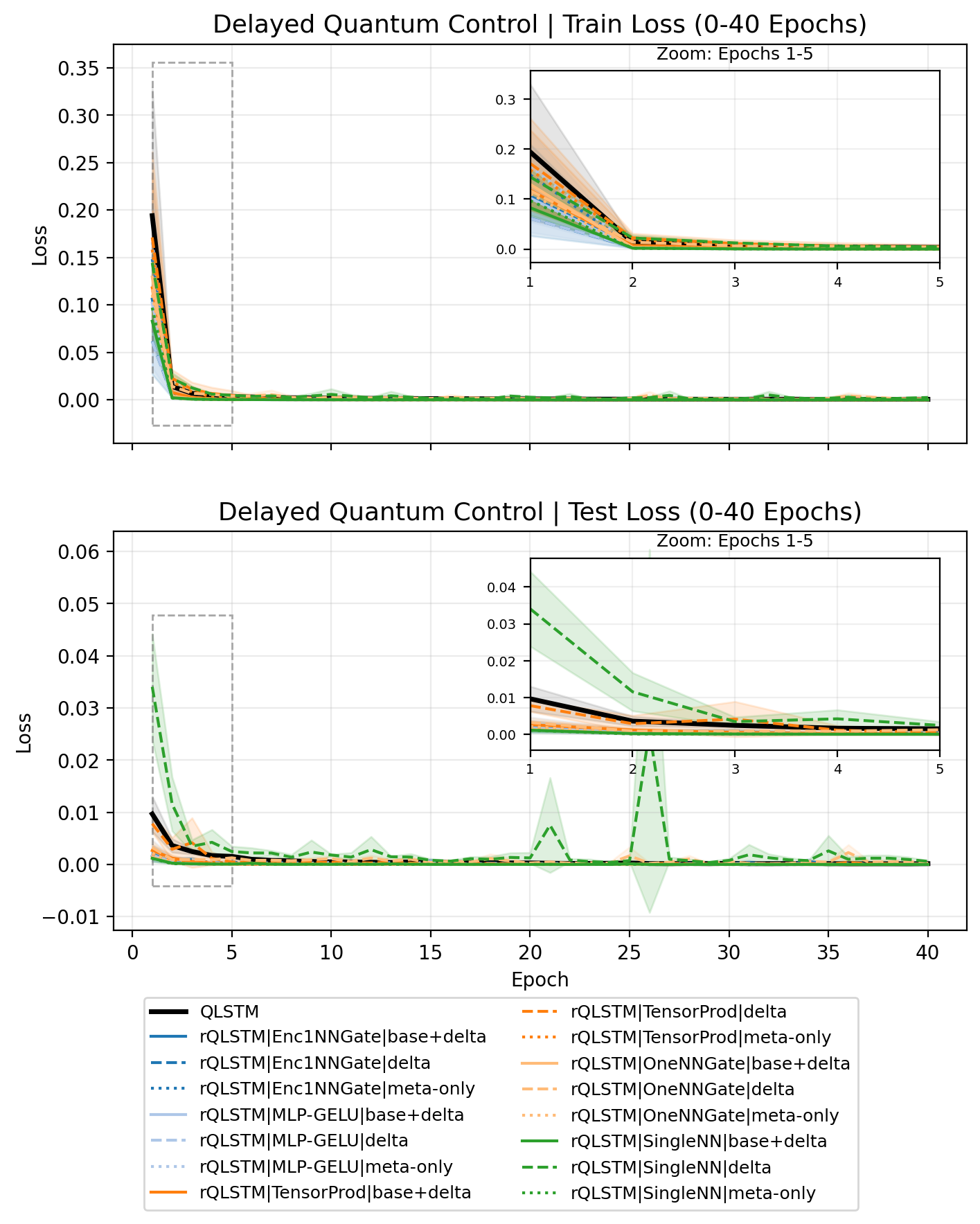}
\caption{\textbf{Train and test loss convergence comparison on the \texttt{delayed\_quantum\_control} dataset with \texttt{seq\_len}=16.} We compare the baseline QLSTM against all rQLSTM variants with different MetaCore designs and recursive rules. Curves show the mean over seeds with shaded regions indicating standard deviation, and the inset zooms into the first five epochs.}
\label{fig:delayed_quantum_control_phase_0_convergence_compare}
\vskip -0.1in
\end{figure}

\figureautorefname{\ref{fig:delayed_quantum_control_phase_0_convergence_compare}} reports the phase-0 train/test convergence results at sequence length $16$, comparing all recursive variants against the QLSTM baseline. As in the previous tasks, most recursive variants drive both training and testing loss to near-zero within only a few epochs, whereas the baseline QLSTM decreases more gradually. The figure also reveals that not all candidate recursive variants are equally stable: one candidate group exhibits noticeably larger fluctuations and occasional spikes in test loss, despite the generally fast optimization trend. This observation further justifies the role of phase 0 as a screening stage, whose purpose is not only to identify fast-converging variants but also to exclude less robust combinations before the full sequence-length study.
\begin{figure}[htbp]
\vskip -0.1in
\centering
\includegraphics[width=1\columnwidth]{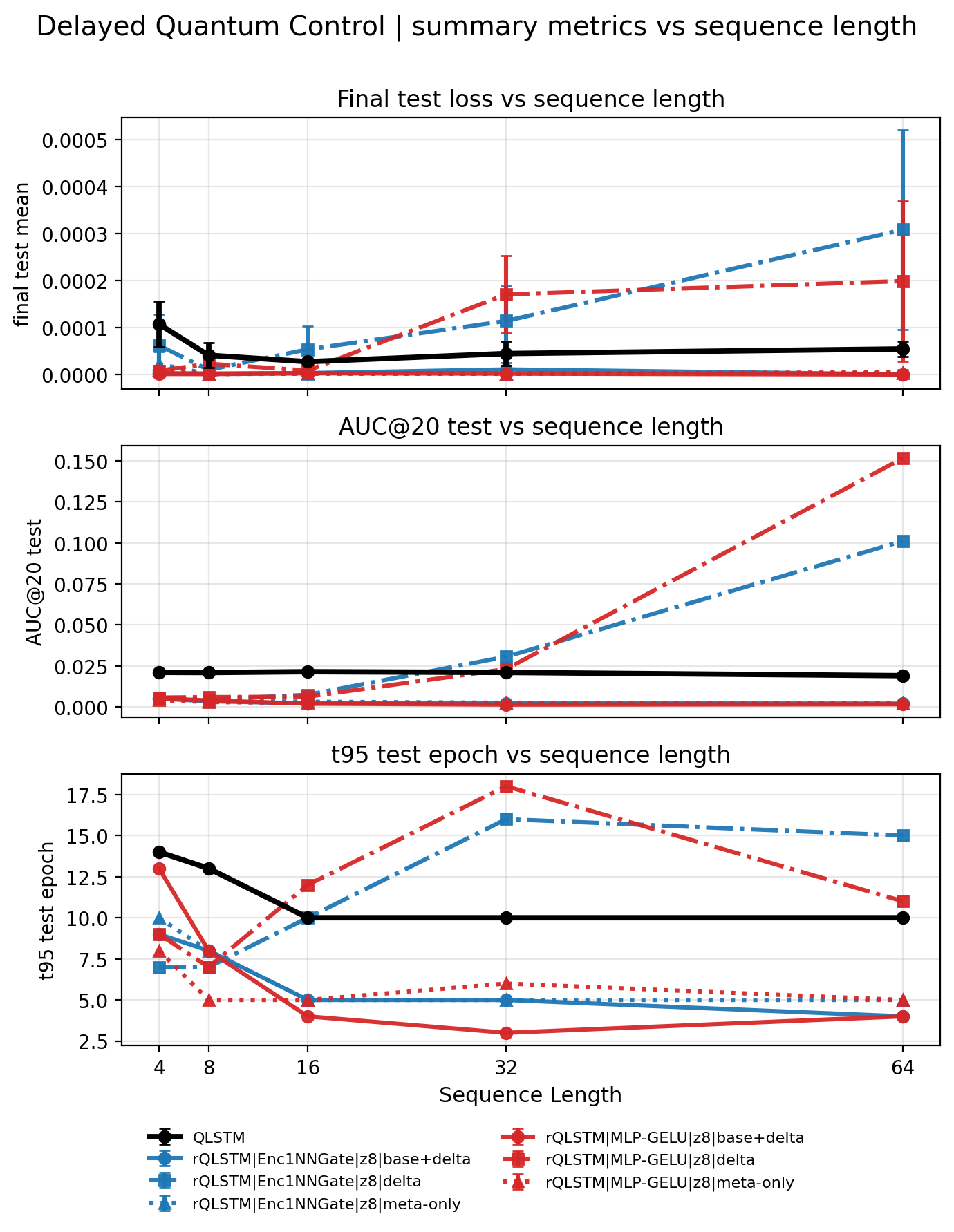}
\caption{\textbf{Summary performance comparison versus sequence length on the \texttt{delayed\_quantum\_control} dataset for sequence lengths 4, 8, 16, 32, and 64.} We compare the baseline QLSTM with selected rQLSTM variants under different MetaCore designs and recursive update rules. The three panels report final test loss, AUC@20 test loss, and $t95$ test epoch, respectively.}
\label{fig:delayed_quantum_control_summary_over_seq_len}
\vskip -0.1in
\end{figure}

\figureautorefname{\ref{fig:delayed_quantum_control_summary_over_seq_len}} summarizes the phase-1 results across sequence lengths $4, 8, 16, 32,$ and $64$ for the retained variants, with the three panels reporting final test loss, AUC@20 test loss, and $t95$ test epoch, respectively. The recursive models remain advantageous in optimization-oriented metrics at short and medium sequence lengths. However, the \texttt{delta} rule becomes noticeably less stable as the sequence length increases: in particular, several \texttt{delta}-based variants exhibit substantially larger AUC@20 and $t95$ at sequence lengths $32$ and $64$. By contrast, the \texttt{base+delta} variants remain stable across sequence lengths, while the \texttt{meta-only} variants are generally more controlled than the corresponding \texttt{delta} settings. Overall, the \texttt{delayed\_quantum\_control} results suggest that recursive QLSTM designs still provide faster learning, but that the more structured \texttt{base+delta} rule offers better robustness across temporal horizons.
\paragraph{NARMA Benchmarks.}
We finally consider the two NARMA tasks, \texttt{narma\_5} and \texttt{narma\_10}, which form a closely related benchmark family and exhibit broadly similar qualitative behavior; they are therefore discussed jointly, with task-specific differences highlighted where relevant.

\begin{figure}[htbp]
\vskip -0.1in
\centering
\includegraphics[width=1\columnwidth]{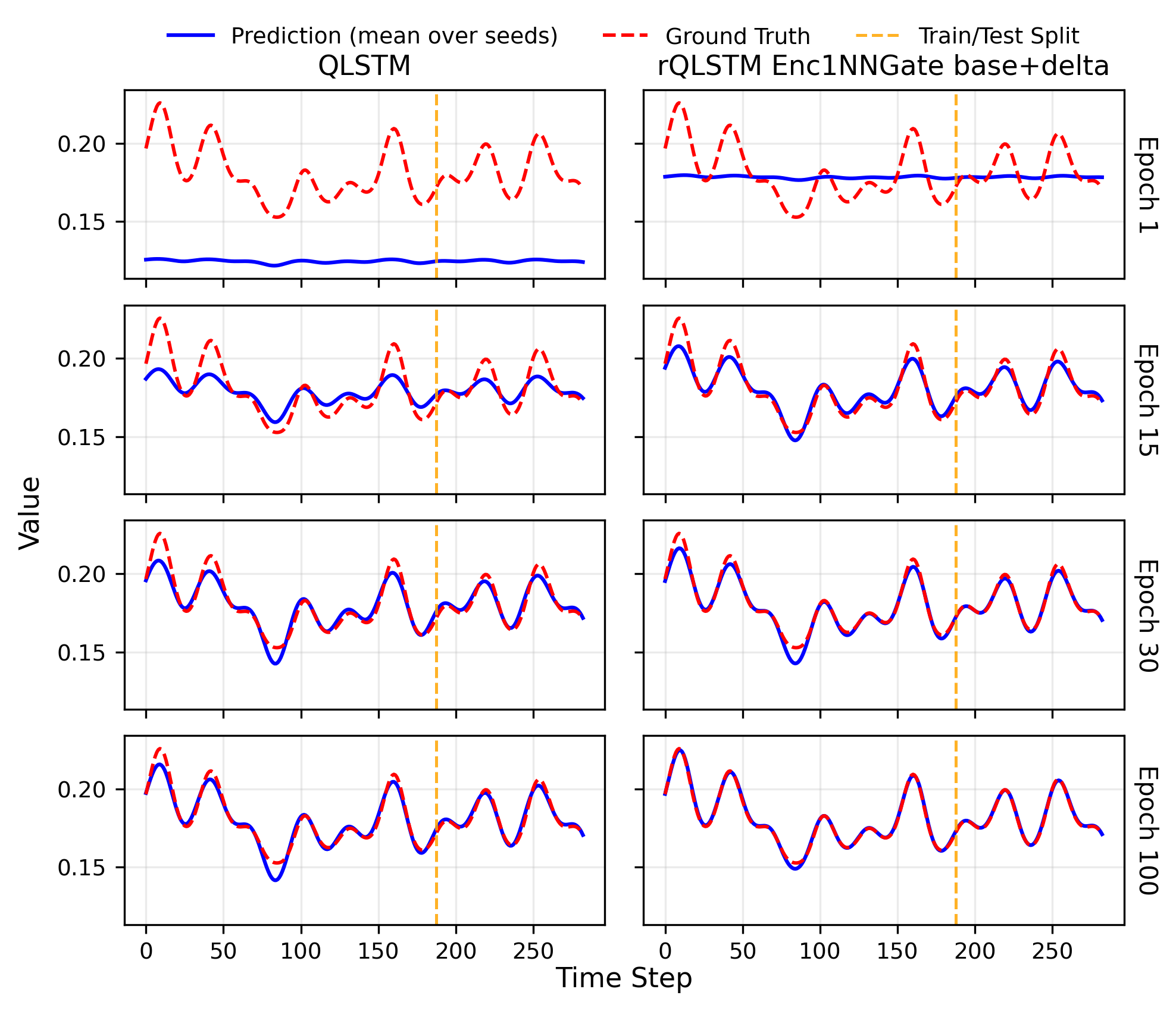}
\caption{\textbf{Epoch-wise prediction comparison between QLSTM and rQLSTM Enc1NNGate base+delta on the \texttt{narma\_5} dataset under the \texttt{seq\_len}=16 setting.} Rows show selected training epochs. Blue curves denote the mean prediction over seeds, red dashed curves denote the ground truth, and the orange dashed line marks the train/test split.}
\label{fig:narma_5_rollout}
\vskip -0.1in
\end{figure}

\begin{figure}[htbp]
\vskip -0.15in
\centering
\includegraphics[width=1\columnwidth]{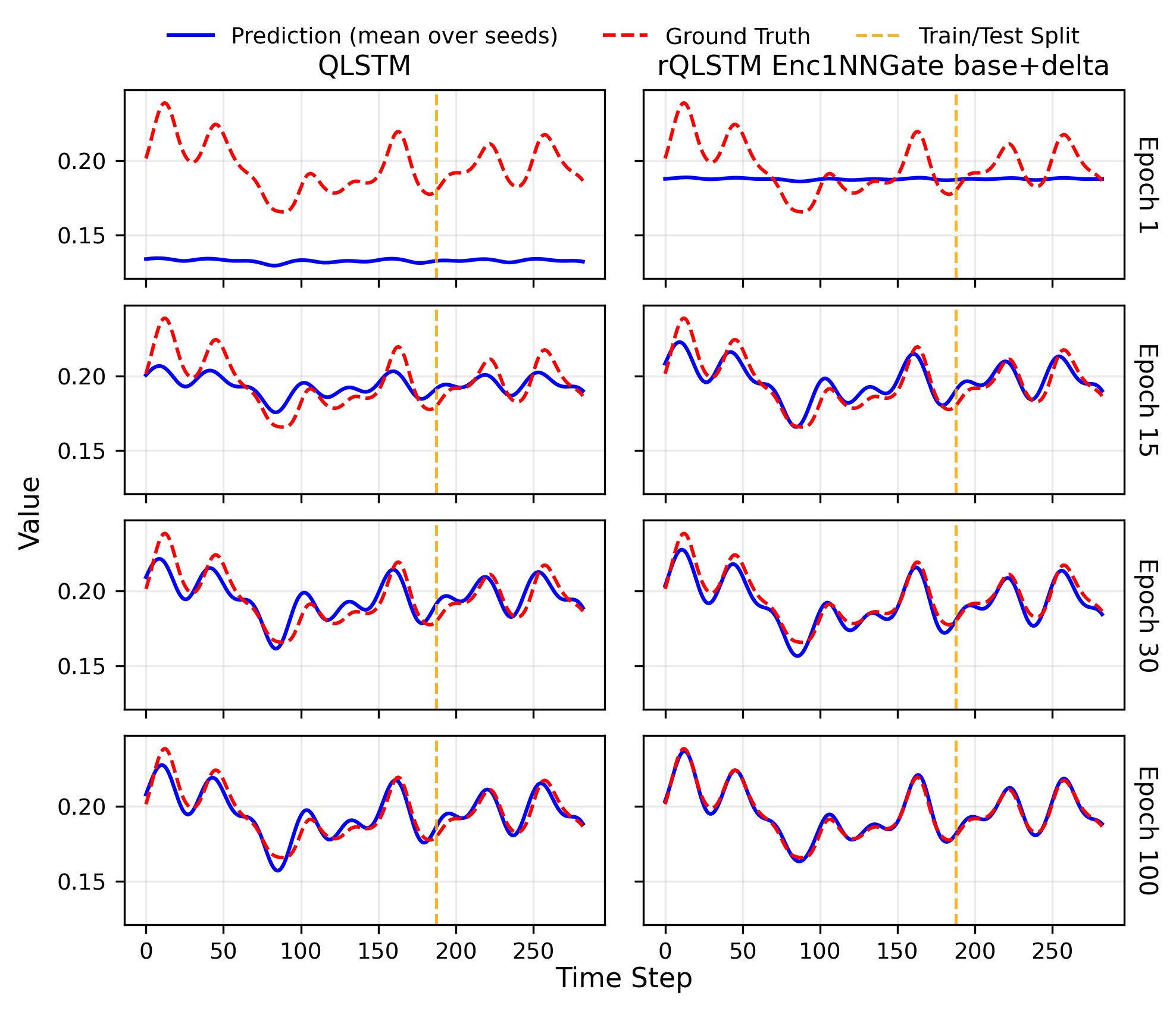}
\caption{\textbf{Epoch-wise prediction comparison between QLSTM and rQLSTM Enc1NNGate base+delta on the \texttt{narma\_10} dataset under the \texttt{seq\_len}=16 setting.} Rows show selected training epochs. Blue curves denote the mean prediction over seeds, red dashed curves denote the ground truth, and the orange dashed line marks the train/test split.}
\label{fig:narma_10_rollout}
\vskip -0.15in
\end{figure}

\figureautorefname{\ref{fig:narma_5_rollout}} and \figureautorefname{\ref{fig:narma_10_rollout}} show the epoch-wise prediction snapshots at sequence length $16$. In both tasks, the representative recursive model \texttt{rQLSTM Enc1NNGate base+delta} exhibits a clear advantage over the standard QLSTM throughout the early and intermediate training stages: up to epoch 15, it is visibly better aligned with the ground-truth trajectory; by epoch 30 the gap narrows but the recursive model remains slightly closer to the target waveform; and only by epoch 100 do the two models become nearly indistinguishable. Thus, even for the NARMA family, where both models eventually achieve strong final predictions, the recursive formulation provides a noticeable advantage in learning speed and intermediate-stage fitting quality.
\begin{figure}[htbp]
\vskip -0.1in
\centering
\includegraphics[width=1\columnwidth]{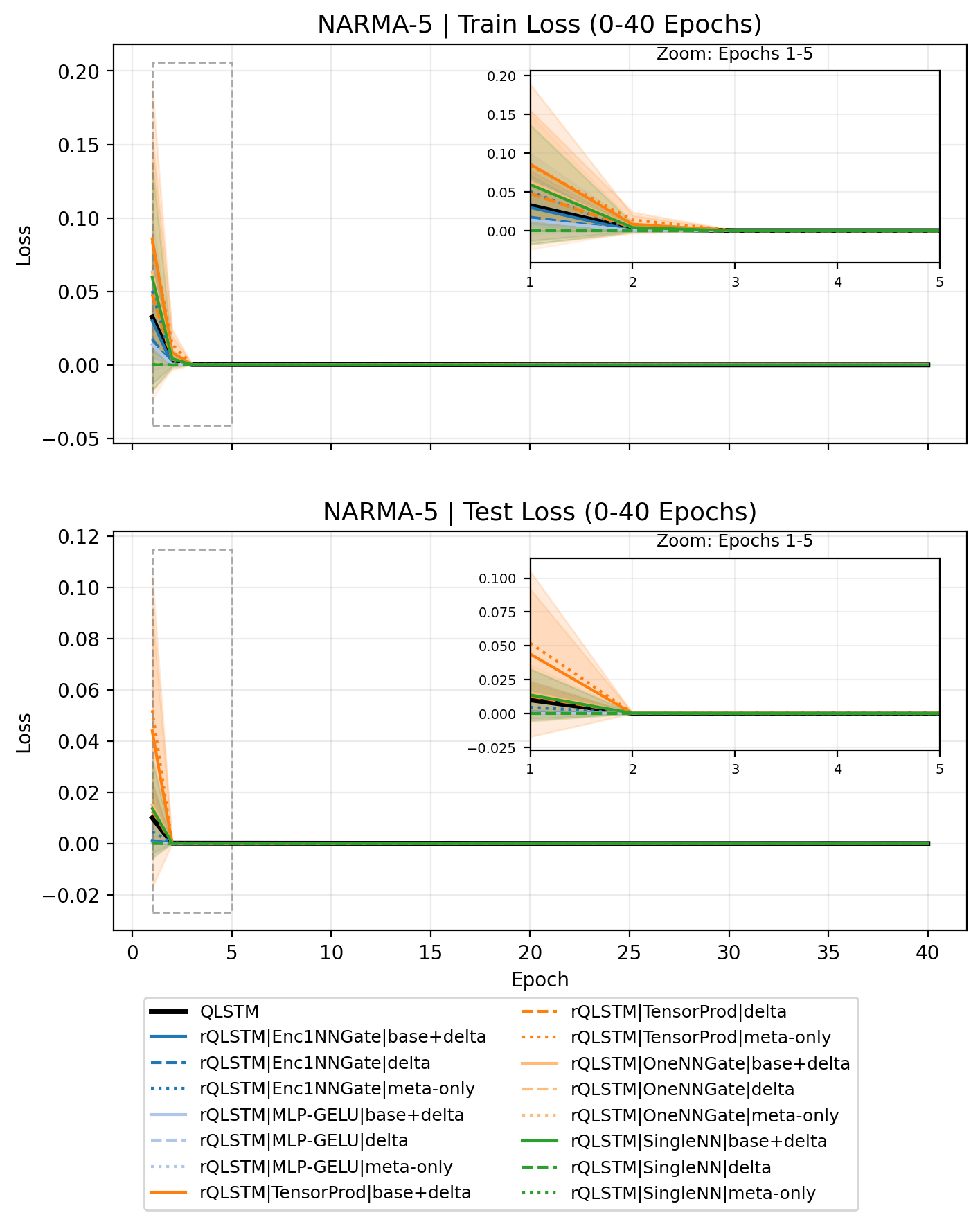}
\caption{\textbf{Train and test loss convergence comparison on the \texttt{narma\_5} dataset with \texttt{seq\_len}=16.} We compare the baseline QLSTM against all rQLSTM variants with different MetaCore designs and recursive rules. Curves show the mean over seeds with shaded regions indicating standard deviation, and the inset zooms into the first five epochs.}
\label{fig:narma_5_phase_0_convergence_compare}
\vskip -0.1in
\end{figure}
\begin{figure}[htbp]
\vskip -0.15in
\centering
\includegraphics[width=1\columnwidth]{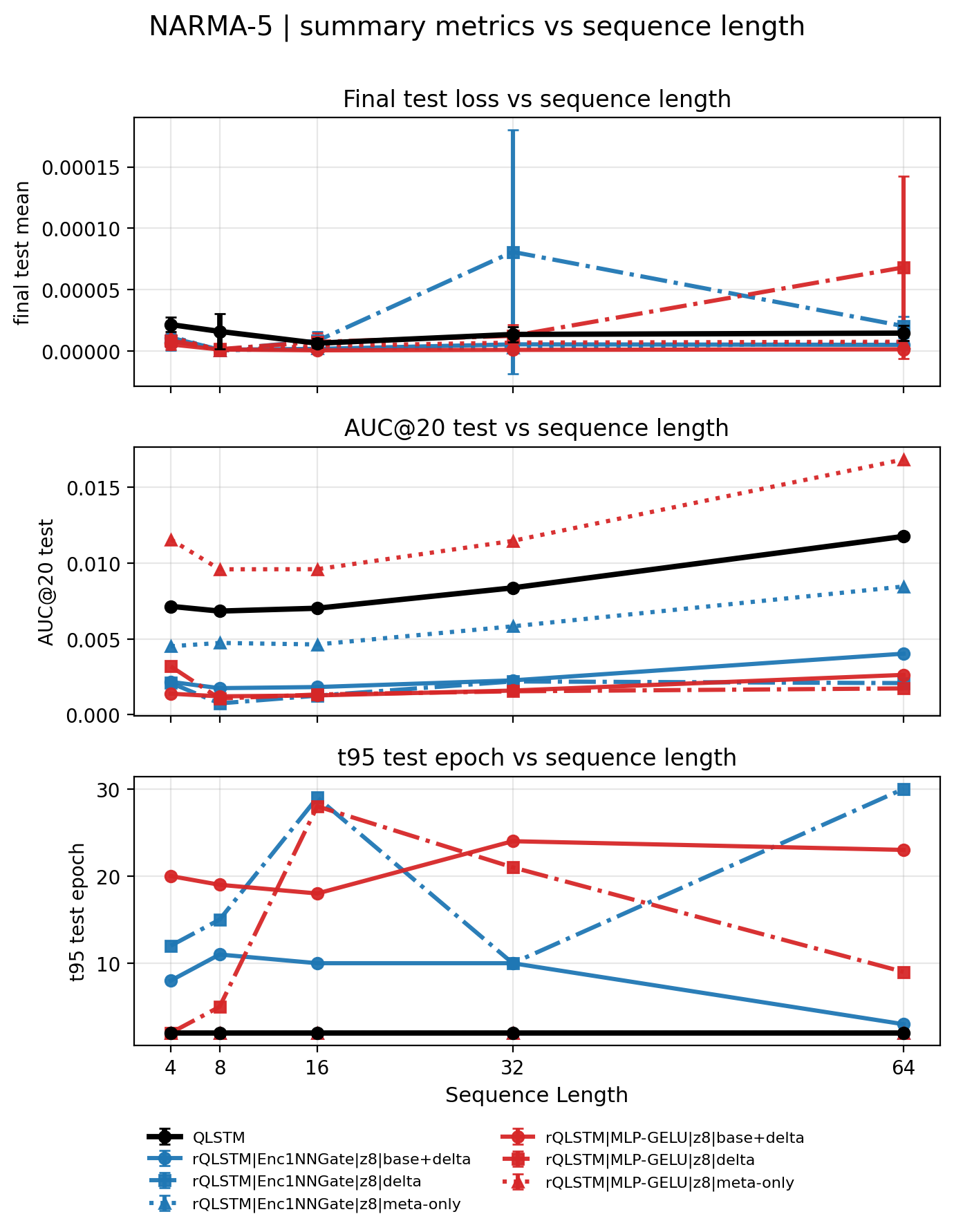}
\caption{\textbf{Summary performance comparison versus sequence length on the \texttt{narma\_5} dataset for sequence lengths 4, 8, 16, 32, and 64.} We compare the baseline QLSTM with selected rQLSTM variants under different MetaCore designs and recursive update rules. The three panels report final test loss, AUC@20 test loss, and $t95$ test epoch, respectively.}
\label{fig:narma_5_summary_over_seq_len}
\vskip -0.15in
\end{figure}

\begin{figure}[htbp]
\vskip -0.15in
\centering
\includegraphics[width=1\columnwidth]{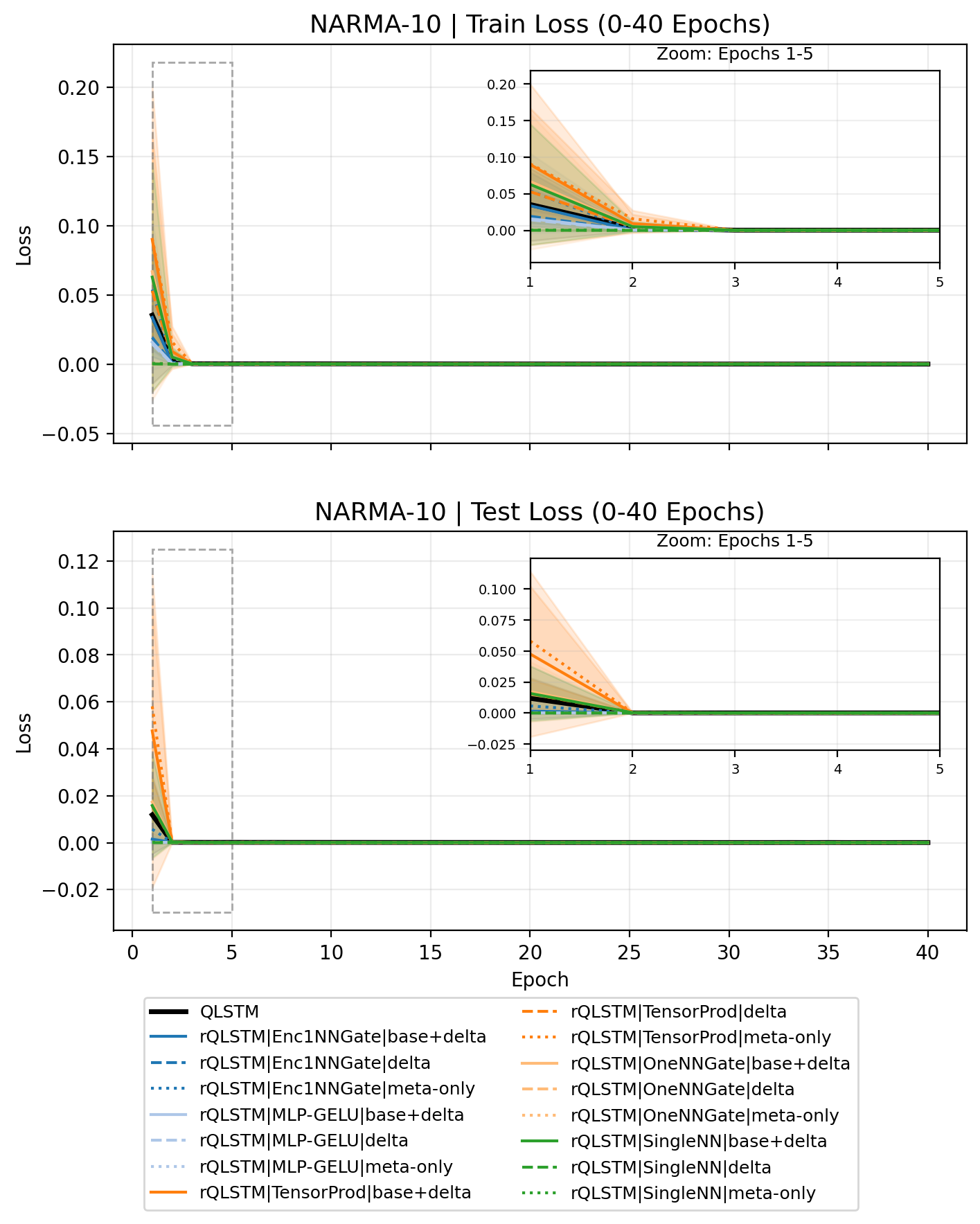}
\caption{\textbf{Train and test loss convergence comparison on the \texttt{narma\_10} dataset with \texttt{seq\_len}=16.} We compare the baseline QLSTM against all rQLSTM variants with different MetaCore designs and recursive rules. Curves show the mean over seeds with shaded regions indicating standard deviation, and the inset zooms into the first five epochs.}
\label{fig:narma_10_phase_0_convergence_compare}
\vskip -0.15in
\end{figure}

\figureautorefname{\ref{fig:narma_5_phase_0_convergence_compare}} and \figureautorefname{\ref{fig:narma_10_phase_0_convergence_compare}} report the phase-0 train/test convergence results at sequence length $16$. 
In both tasks, nearly all recursive variants reduce both training and testing loss to near-zero within a few epochs, while the baseline QLSTM also converges quickly and remains relatively competitive on \texttt{narma\_5} and only moderately slower on the slightly more demanding \texttt{narma\_10}. 
Phase 0 on the NARMA family is therefore less discriminative than in the earlier tasks, since the optimization problem is already fairly easy at this horizon. 
Nevertheless, the recursive family still shows      a consistent early-convergence advantage, and the screening also reveals that not all candidate MetaCore families are equally competitive: the \texttt{TensorProd} variants in particular appear less favorable on both tasks and are therefore excluded from the subsequent phase-1 study.
\begin{figure}[htbp]
\vskip -0.15in
\centering
\includegraphics[width=1\columnwidth]{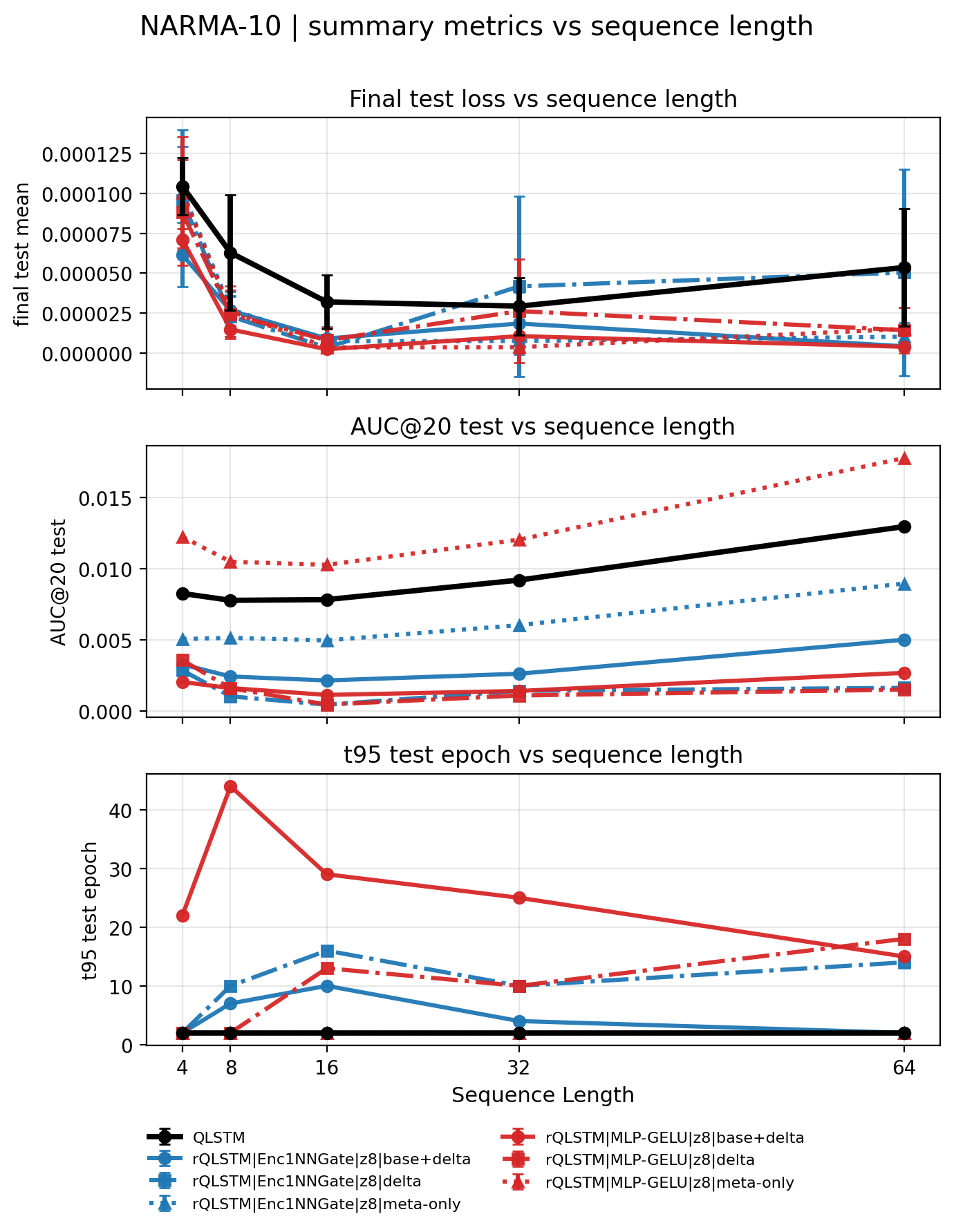}
\caption{\textbf{Summary performance comparison versus sequence length on the \texttt{narma\_10} dataset for sequence lengths 4, 8, 16, 32, and 64.} We compare the baseline QLSTM with selected rQLSTM variants under different MetaCore designs and recursive update rules. The three panels report final test loss, AUC@20 test loss, and $t95$ test epoch, respectively.}
\label{fig:narma_10_summary_over_seq_len}
\vskip -0.15in
\end{figure}

\figureautorefname{\ref{fig:narma_5_summary_over_seq_len}} and \figureautorefname{\ref{fig:narma_10_summary_over_seq_len}} summarize the phase-1 results across sequence lengths $4, 8, 16, 32,$ and $64$ for the retained variants, with each set of three panels reporting final test loss, AUC@20 test loss, and $t95$ test epoch, respectively. 
For both tasks, once final accuracy has largely saturated for most models, the main remaining distinction lies in training efficiency and robustness rather than in absolute end-point accuracy alone. 
The summary metrics reveal consistent differences among the retained rules across both tasks: the \texttt{meta-only} variants, especially under \texttt{MLP-GELU}, tend to exhibit relatively worse AUC@20, indicating weaker early-stage error reduction; the \texttt{delta} variants show more noticeable fluctuation in final test loss across sequence lengths, which is particularly pronounced on the more demanding \texttt{narma\_10}; and the \texttt{base+delta} variants are more favorable in convergence speed, with some settings showing a tendency toward smaller $t95$ as the sequence length increases. 
Overall, the NARMA results again support the effectiveness of recursive QLSTM designs, while showing that \texttt{base+delta} provides the most balanced trade-off among final accuracy, early-training efficiency, and convergence robustness.

Overall, the experimental results across all five tasks show that the main advantage of the proposed recursive QLSTM family lies in \emph{learning efficiency} rather than only in final end-point accuracy. The epoch-wise prediction snapshots show that the representative rQLSTM model recovers the target temporal structure earlier than the standard QLSTM, with the advantage clearly visible at intermediate training stages on most tasks and only gradually narrowing at later epochs. The phase-0 convergence study at sequence length $16$ further supports this trend on \texttt{bessel\_j2}, \texttt{damped\_shm}, and \texttt{delayed\_quantum\_control}, while phase 0 on the NARMA benchmarks is less discriminative since the baseline QLSTM itself converges quickly. The subsequent phase-1 sequence-length study confirms that the recursive family generally remains competitive or superior in final test loss while showing clearer gains in AUC@20 and $t95$ across temporal horizons, although these gains are not uniform across all rules: in particular, \texttt{delta}-based variants degrade noticeably at longer sequence lengths on \texttt{delayed\_quantum\_control} and \texttt{narma\_10}, and \texttt{meta-only} variants are comparatively weaker in AUC@20, especially on the NARMA benchmarks. Among the retained rules, \texttt{base+delta} provides the most balanced and robust behavior overall. Taken together, these results suggest that recursion itself is the primary source of improvement, while the specific MetaCore and update-rule choices mainly govern the trade-off between stability, robustness, and convergence speed.
\section{Theoretical Discussion}

The \texttt{base+delta} rule $\Theta_t^a=\bar\Theta^a+\Delta_t^a$ and $a\in\mathcal{A}=\{i,f,g,o\}$ can be viewed as a residual perturbation of a learned base QLSTM. 

We analyze the noiseless expectation-value VQC map $u \mapsto V(u;\Theta) \in \mathbb{R}$ at parameter $\Theta$. Since the circuit in Sec.~\ref{Sec: QNN} is composed of smooth parameterized $R_Y$ rotations and fixed entangling layers, $\Theta \mapsto V(u;\Theta)$ is smooth on bounded parameter regions. Hence, for each gate $a\in\mathcal G$, we have Taylor expansion at $\bar\Theta^a$
\begin{equation}
V(u_t;\bar\Theta^a+\Delta_t^a) = V(u_t;\bar\Theta^a) + \langle J_t^a, \Delta_t^a \rangle + R_t^a,
\end{equation}
with $J_t^a:=D_\Theta V(u_t;\bar\Theta^a)$ and $ \|R_t^a\|\le C_V\|\Delta_t^a\|_F^2$ on any compact region containing $\bar\Theta^a + \Delta_t^a$. Thus the \texttt{base+delta} gates are first-order context-dependent corrections of the base QLSTM gates.

Let $s_t=(h_t,c_t)$ and write the recurrent update as
$s_t=F_{\bar{\Theta}+\Delta_t}(s_{t-1},x_t)$, with $s_t^{(0)}$ denoting the
trajectory of the base QLSTM. By smoothness of the VQC and the Lipschitzness of
the gate nonlinearities such as $\sigma$ and $\tanh$, there exists $C>0$ on bounded trajectories such that
\begin{equation}
\|F_{\bar{\Theta}+\Delta_t}(s,x)-F_{\bar{\Theta}}(s,x)\|
\le
C\sum_{a\in\mathcal{A}}\|\Delta_t^a\|_F
+
O\!\left(\sum_{a\in\mathcal{A}}\|\Delta_t^a\|_F^2\right).
\end{equation}
Assume the base QLSTM is \textbf{locally contractive}, \textit{i.e.},
\begin{equation}
\|F_{\bar{\Theta}}(s,x)-F_{\bar{\Theta}}(s',x)\|
\le
\rho\|s-s'\|,
\qquad 0\le \rho<1,
\end{equation}
then the deviation from the base trajectory satisfies
\begin{align}
\|s_t-s_t^{(0)}\|
&\le
\rho^t\|s_0-s_0^{(0)}\|
+
C\sum_{\tau=1}^{t}\rho^{t-\tau}
\sum_{a\in\mathcal{A}}\|\Delta_\tau^a\|_F \nonumber\\
&\quad+
O\!\left(
\sum_{\tau=1}^{t}\rho^{t-\tau}
\sum_{a\in\mathcal{A}}\|\Delta_\tau^a\|_F^2
\right).
\end{align}
\begin{IEEEproof}[Derivation]
Let $d_t=\|s_t-s_t^{(0)}\|$. Adding and subtracting
$F_{\bar{\Theta}}(s_{t-1},x_t)$ gives
\begin{align}
d_t
&\le
\|F_{\bar{\Theta}+\Delta_t}(s_{t-1},x_t)
-
F_{\bar{\Theta}}(s_{t-1},x_t)\| \nonumber\\
&\quad+
\|F_{\bar{\Theta}}(s_{t-1},x_t)
-
F_{\bar{\Theta}}(s_{t-1}^{(0)},x_t)\|.
\end{align}
Using the parameter-perturbation bound and the local contraction of the base
QLSTM, we obtain
\begin{equation}
d_t
\le
\rho d_{t-1}
+
C\sum_{a\in\mathcal{A}}\|\Delta_t^a\|_F
+
O\!\left(
\sum_{a\in\mathcal{A}}\|\Delta_t^a\|_F^2
\right).
\end{equation}
Unrolling this recursion yields
\begin{align}
d_t
&\le
\rho^t d_0
+
C\sum_{\tau=1}^{t}\rho^{t-\tau}
\sum_{a\in\mathcal{A}}\|\Delta_\tau^a\|_F \nonumber\\
&\quad+
O\!\left(
\sum_{\tau=1}^{t}\rho^{t-\tau}
\sum_{a\in\mathcal{A}}\|\Delta_\tau^a\|_F^2
\right).
\end{align}
\end{IEEEproof}
Consequently, bounded instantaneous residuals imply bounded deviation from the
learned base QLSTM; for example, if
$\sum_a\|\Delta_t^a\|_F\le \epsilon$ and the initial states coincide, then
$\limsup_t\|s_t-s_t^{(0)}\|\le C\epsilon/(1-\rho)+O(\epsilon^2)$.

This explains the stability advantage of the proposed design. A \texttt{meta-only}
rule must generate the entire VQC from the current context and lacks an explicit
anchor. A pure recursive-delta rule \texttt{delta},
$\Theta_t^a=\Theta_{t-1}^a+\Delta_t^a$, accumulates residuals over time, so
zero-mean residual noise can induce random-walk variance and biased residuals
can cause linear drift. In contrast, \texttt{base+delta} uses only instantaneous
residuals around a fixed learned circuit. The shared encoder further reduces
variance by learning common recurrent features, while the gate-specific heads
allow the input, forget, candidate, and output gates to apply different
corrections. Thus the architecture provides a favorable stability--adaptivity
trade-off without claiming strict expressivity dominance over an equivalent
block-partitioned MLP implementation.

\section{Conclusion}
This paper proposed Recursive QLSTM, a MetaCore-based extension of QLSTM that dynamically adapts VQC parameters from the recurrent context. Experiments on five time-series benchmarks across multiple sequence lengths show that recursive parameter adaptation mainly improves learning efficiency, yielding faster early-stage error reduction, lower AUC@20, and smaller $t95$ while maintaining competitive final test loss. Among the evaluated designs, the base+delta rule with the shared-encoder gate-specific MetaCore provides the most stable and balanced performance. The theoretical analysis further interprets this rule as a bounded residual perturbation around a learned base QLSTM, explaining its stability--adaptivity trade-off. These results suggest that recursive VQC parameter adaptation is an effective mechanism for improving quantum recurrent sequence models.

\clearpage
\bibliographystyle{IEEEtran}
\bibliography{bib/qml_examples,bib/classical_sequence,bib/qml_foundations,bib/quantum_recurrent,bib/dynamic_recursive,bib/quantum_architecture_search,bib/qc}

\end{document}